\documentclass[aps,prc,twocolumn,showpacs,preprintnumbers,amsmath,amssymb,superscriptaddress]{revtex4}
\usepackage{graphicx,amsmath,color}
\usepackage{latexsym}

\begin{document}

\title{Observation of a narrow structure in $^1H(\gamma,K_S^0)X$ \\
 via interference with $\phi$-meson production} 
\newcommand*{\ODU}{Old Dominion University, Norfolk, Virginia 23529}
\affiliation{\ODU}
\newcommand*{\BOCHUM}{Institut f\"ur Theoretische Physik II,
Ruhr--Universit\"at Bochum, D--44780 Bochum, Germany}
\newcommand*{\PNPI}{Petersburg Nuclear Physics
Institute, Gatchina, St.\ Petersburg 188300, Russia}
\newcommand*{\KYUNGPOOK} {Kyungpook National University, 702-701, Daegu, Republic of Korea}
\newcommand*{\INR}{Institute for Nuclear Research, 117312, Moscow, Russia}
\newcommand*{\GWU}{The George Washington University, Washington, DC 20052}
\newcommand*{\CUA}{Catholic University of America, Washington, DC 20064} 
\newcommand*{\JLAB}{Thomas Jefferson National Accelerator Facility, Newport News, Virginia 23606}

\author {M.J.~Amaryan}
\email{mamaryan@odu.edu}
\thanks{Corresponding author.}
\affiliation{\ODU}
\author {G.~Gavalian}
\affiliation{\ODU}
\author {C.Nepali}
\affiliation{\ODU}
\author {M.V.~Polyakov}
\affiliation{\BOCHUM} \affiliation{\PNPI}
\author {Ya.~Azimov}
\affiliation{\PNPI}
\author {W.J.~Briscoe}
\affiliation{\GWU}
\author {G.E.~Dodge}
\affiliation{\ODU}
\author {C.E.~Hyde}
\affiliation{\ODU}
\author{F.~Klein}
\affiliation{\CUA}
\author{V.~Kuznetsov}
\affiliation{\KYUNGPOOK} \affiliation{\INR}
\author{I.~Strakovsky}
\affiliation{\GWU}
\author{J.~Zhang}
\affiliation{\JLAB}

\date{\today}

\begin{abstract}

We report observation of a narrow peak structure at $\sim$1.54~GeV
with a Gaussian width $\sigma$~=~6~MeV in the missing mass of $K_S$ in the reaction
$\gamma+p \to p K_S  K_L$. The observed structure may be due to
the interference between a strange (or anti-strange) baryon
resonance in the $pK_L$ system and the $\phi(K_SK_L)$ photoproduction
leading to the same final state. The statistical
significance of the observed excess of events estimated as
the log likelihood ratio of the resonant signal+background
hypothesis and the $\phi$-production based background-only
hypothesis  corresponds to $5.3\sigma$.
\end{abstract}


\maketitle

\section{ Introduction}

The Non-Relativistic Constituent Quark Model (NRCQM) describes mesons and baryons
as $q\bar q$ pairs and  $3q$ configurations respectively. Proposed originally to describe classification
of light mesons and baryons consisting of u, d and s quarks, NRCQM appears to be very successful.
In particular, in the baryon sector, which is more relevant to this
study, it unifies all known  light baryons in terms of two 
irreducible representations of SU(3) symmetry: spin 1/2 baryons belonging to  the octet ($\bf 8$)  and their excited 
states together with the $\Omega^-$ hyperon, all spin 3/2 states,
belonging to the decuplet ($\bf 10$). 
Furthermore, states with different isospin projections, but the same hypercharge, form isospin multiplets. 

However the NRCQM is a phenomenological model. It is not  derived from 
first principles of quantum chromodynamics (QCD), 
the fundamental theory of strong interactions, and therefore the existence of other states beyond its 
limits can not be excluded. Among such states are hybrids, glueballs, and multiquark states. 
The observation of these new QCD configurations, or understanding the reason why they are not realized in  nature, will 
help to obtain an important insight into the underlying
dynamics of strong interactions
and properties of QCD in the
non-perturbative regime.

Discussions about the multiquark states go back to the early days of the
quark model and unsuccessful experimental efforts to observe such 
configurations span  decades. 
However, recently there appeared a striking prediction of the Chiral
Quark Soliton model~\cite{DIAKON} for an entire new family of 5-quark (pentaquark)
states that belong to the $\bf \bar {10}$ (anti-decuplet) representation
of SU(3) symmetry, creating a new wave of excitement in the field of hadronic physics.
In particular an explicitly exotic pentaquark state, with minimal
quark content  $uudd\bar s$,  
lying on the apex of the new representation of 
SU(3) symmetry and called now $\Theta^+$, was predicted to have
mass $M_{\Theta^+}=1.53$~GeV and narrow width $\Gamma <15$~MeV. 
From an experimental point of view this
excitement was due to the narrow width of the predicted 
pentaquark state, which would make its observation 
much easier, due to its simple decay mode to $K^+n$ or
$K^0p$, and finally due to its relatively low mass, which makes its production possible
 at many  experimental facilities.
 
Inspired by this prediction, the first experimental results of the
observation of $\Theta^+$ were obtained and reported by 
the SPring-8 Collaboration in a low energy photoproduction experiment~\cite{nakano} and
independently by the DIANA Collaboration~\cite{DIANA} in a formation reaction
with a low energy kaon beam. Subsequently positive claims followed by
the CLAS~\cite{CLASD},~\cite{CLASH}, SAPHIR~\cite{SAPHIR}, HERMES~\cite{HERMES}, 
ZEUS~\cite{ZEUS} and SVD~\cite{SVD} Collaborations. In parallel,
negative results were reported by several groups: HERA-B~\cite{HERAB},
HyperCP~\cite{HyperCP},  BES~\cite{BES}, ALEPH~\cite{ALEPH}, and
BABAR~\cite{BABAR}.

The common feature of most experimental results was that they were
reported out of non dedicated experiments.  It was not until 2004 that the CLAS Collaboration 
performed dedicated high statistics photoproduction experiments both on deuterium and hydrogen targets.
It was found, first of all, that the previous measurement on the deuterium
target by CLAS~\cite{CLASD} was not reproduced in the new measurement,
despite an
order of magnitude higher statistics~\cite{CLASg10}.
It is now understood ~\cite{CLASg10} that the level of background in the
first paper was underestimated and therefore the observed signal was statistically not
so significant.
The search for $\Theta^+$  in the high
statistics CLAS measurement of the reaction $\gamma + p\rightarrow \bar K^0K^+n$ 
in CLAS~\cite{CLAS1} was negative. 
This is the same channel in which the SAPHIR collaboration reported a
positive result~\cite{SAPHIR}. 
It is worthwhile to mention that there is a significant difference in the
geometric acceptance of  CLAS and SAPHIR in the forward direction:
on the order of a few percent for the $\pi^+\pi^-K^+$ triple coincidence
in CLAS and almost full coverage in the forward angles for SAPHIR.
This will affect the sensitivity if the  $\Theta^+$ production is strongly forward peaked.
In addition the search for the  $\Theta^+$ was
performed for the first time  in the reaction $\gamma + p \rightarrow
\bar K^0 K^0p$, which also did not result in a $\Theta^+$ signal~\cite{CLAS1}.
 Before publication of the high statistics CLAS data, the experimental
situation was considered to be uncertain. The high statistics CLAS
publications~\cite{CLASg10,CLAS1}
lowered the confidence for the existence of the $\Theta^+$ pentaquark~\cite{pdg},
although  the  papers themselves quote only upper limits on the 
$\Theta^+$ photoproduction cross section, estimated to be on the
 order of a few nb. Detailed reviews of the experimental situation can be
found in~\cite{hicks, burkert}.  Critical comparison of positive
and negative results was presented in~\cite{AGS}.
Meanwhile  the SPring-8 Collaboration published a new paper~\cite{nakano2},
where they reproduced their previous result with increased
statistics on a deuteron target, and the DIANA
Collaboration confirmed 
their previous result with increased statistics in renewed
analyses~\cite{diana2, diana3}.

\begin{figure}[htb]
 \includegraphics[width=3.2in]{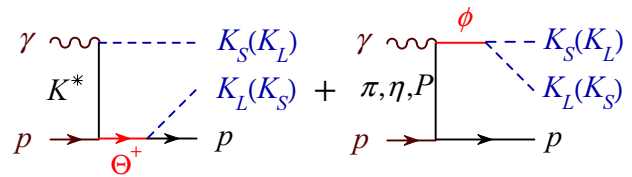}
\caption{(Color online). Two different subprocesses in the reaction
$\gamma p \rightarrow  pK_SK_L$ that can lead to the same final state:
  $\Theta^+(pK^0)$ production
  (left) and $\phi$-meson production (right).}
\label{fig:diagram}
\end{figure}

The analysis reported here was performed  in an attempt to increase the
experimental sensitivity of the CLAS setup to a small $\Theta^+$ signal.
One possible way to do so, is to exploit quantum mechanical
interference to enhance the small amplitude of
 the $\Theta^+$ by some other resonance with a strong production cross section leading to the same final state. 
 Numerous examples of how interference helps to enhance the faint signal of one resonance by a stronger signal of another resonance are presented in a recent review by Azimov~\cite{azimov}.
  Such a possibility for the search of $\Theta^+$ can be realized in the reaction  $\gamma p \rightarrow pK_SK_L$ 
  where, as was proposed in~\cite{amarian}, one can use
  photoproduction of the $\phi(K_SK_L)$ meson to enhance a baryon resonance in either the $pK_{S}$ or $pK_{L}$ system. The two processes leading to the same $pK_SK_L$ final state are shown in 
Fig.~\ref{fig:diagram}. Since both $\gamma p \to p\phi \to pK_SK_L$ and 
$\gamma p \to \Theta^+ \bar {K}^0 \to pK_SK_L$ reactions have the same
final state, quantum mechanically they must interfere. As a result of
the interference, the small amplitude of a  possible $\Theta^+$ (or
any other baryon with a similar decay mode) production would be
multiplied by the large $\phi$ production amplitude, 
thus increasing sensitivity to a possible signal of the strange (or anti-strange) baryon. 

\section{Experiment}

The present study is based on  the same data set collected in 2004 (g11a run period) using the CLAS
detector at TJNAF  and analyzed previously~\cite{CLAS1}.
  The experiment was performed using a photon beam produced through
  bremsstrahlung from a 4.02~GeV initial electron beam from the Continuous Electron Beam Accelerator Facility (CEBAF) at Jefferson Lab.

A scintillator hodoscope system, combined with a dipole magnet, was
used to tag the photon energy in the range of $0.8$ to $3.8~$GeV, with
a resolution of $0.1\%$ of the incident electron energy.
A different set of scintillators are used for the timing measurement.
Surrounding the target segmented (for each sector) scintillator counters were placed for triggering the event.
The CLAS detector is described in detail elsewhere~\cite{mecking}.

In this experiment the photon beam was incident on a 40~cm long
liquid hydrogen target, centered 10~cm upstream from  the center of the
CLAS detector. Particles from the reaction were detected in the CLAS
detector, consisting of six equal sectors, equipped with
time-of-flight  scintillator counters, Electromagnetic
Calorimeters, Drift Chambers and {\v C}erenkov Counters, covering nearly
$4\pi$ solid angle. The  drift chambers consisted of three layers, each layer having
two sub-layers. The second layer was placed inside of a toroidal
magnetic field, used to bend the trajectories of the
charged particles in order to measure their momenta. The momentum resolution of
the CLAS detector is momentum dependent and on average is on the order
of $\Delta P / P \sim0.5\%$.
The charged particle identification is based on simultaneous measurement of their momenta and the time-of-flight.
The CLAS standard particle identification scheme is used to select
charged particles  in the final state. 
The  photon beam energy correction and charged particle momentum correction are based on the code  developed
 by the  g11 run group and used in the previous analysis~\cite{CLAS1}.
 The raw data used in this analysis were processed in the same
 way as in~\cite{CLAS1},
including corrections for the energy loss of charged particles in the target,
uncertainties in the magnetic field, and misalignments of drift chambers.

\section {Analysis}
\subsection{Event Selection and Reconstruction of Final State}
Events for this analysis are selected requiring at least three charged tracks in the final state
identified as a proton, $\pi^+$ and $\pi^-$.  
The initial photon is chosen to be within 1~ns of the start time
defined by the start counter, and it was required to have only one hit in
the tagger within 1.5~ns of the start time. 
 The $K_S$ is reconstructed in the invariant mass of two pions and
 $K_L$ in the missing mass of detected particles,
 $M(K_L)^2=M_X(pK_S)^2=(P_{\gamma}+P_t-P_{K_S}-P_p)^2$, where $P_{i}$
 are four momenta of the photon, target proton, $K_S$ and final state
 proton. The search for a resonance in the $KN$ system  can be done either in the invariant mass of the proton and $K_S$ or in the missing mass of $K_S$. 
To identify  the reconstructed $K_S$ and the final $KN$ state with good mass resolution and acceptable signal to background ratio
the following cuts were implemented (hereinafter referred to as vertex cuts):
\begin{itemize}

\item the proton track must come within 2~cm  of the photon beamline; the
mid-point of the shortest line between the proton track and the photon beam line is
called the primary vertex.

\item The distance of closest approach of the two pion tracks must be less than 1.5~cm;
the mid-point of the shortest line between the two pion tracks is called the decay vertex.

\item $\cos\theta_c > 0.96$; the collinearity angle, $\theta_c$, is the angle between
the line connecting the primary and decay vertices and the direction of the three-momentum
of the $K_S$ reconstructed as the sum of the two pion momenta.

\end{itemize}

The impact of the vertex and particle ID cuts are presented in Fig. 2.
In this figures, all mass distributions in the left column are without,
and in the right column with, the vertex cuts described above.

\begin{figure}[htb]
\includegraphics[width=1.6in] {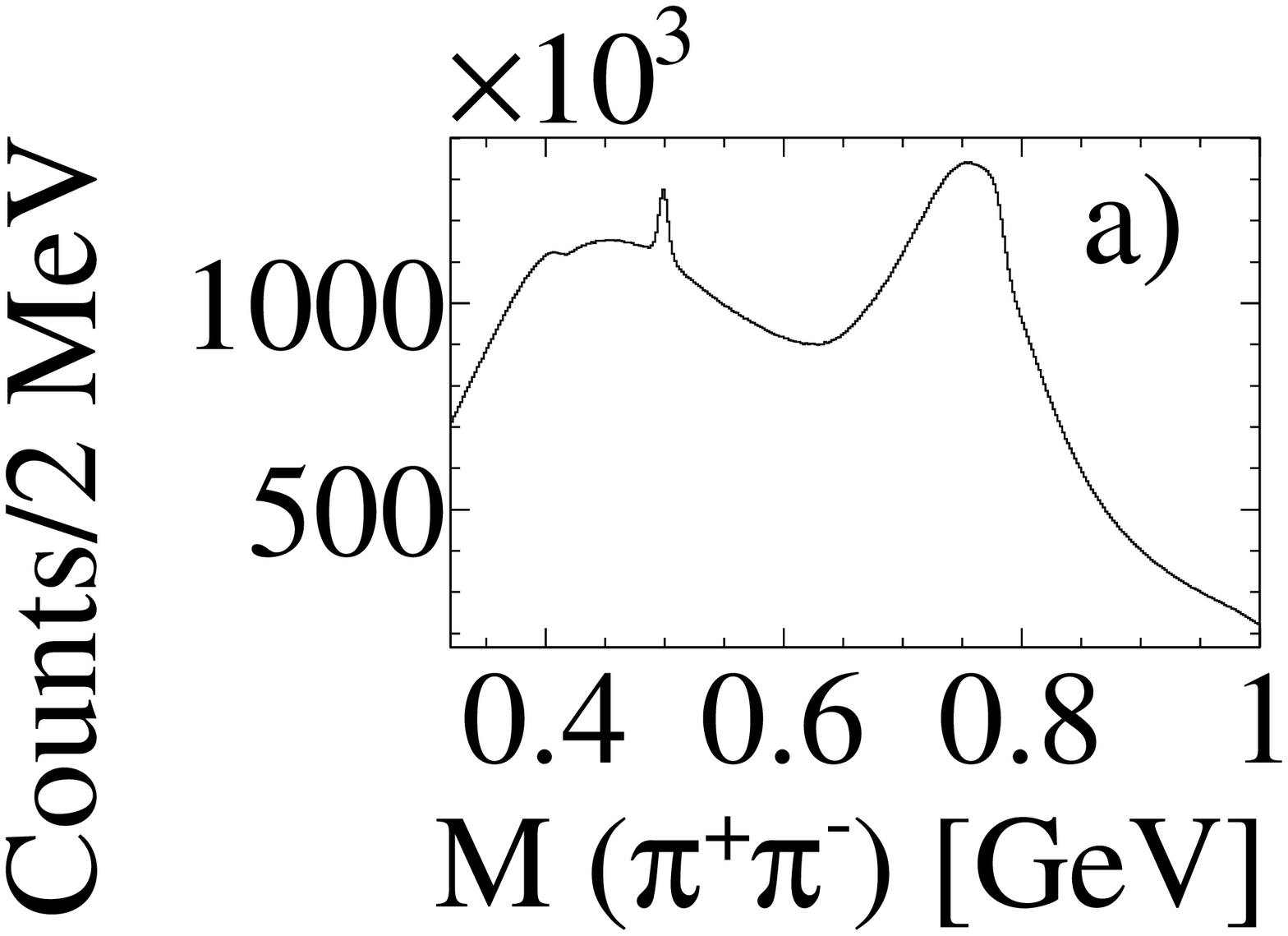}
\includegraphics[width=1.6in] {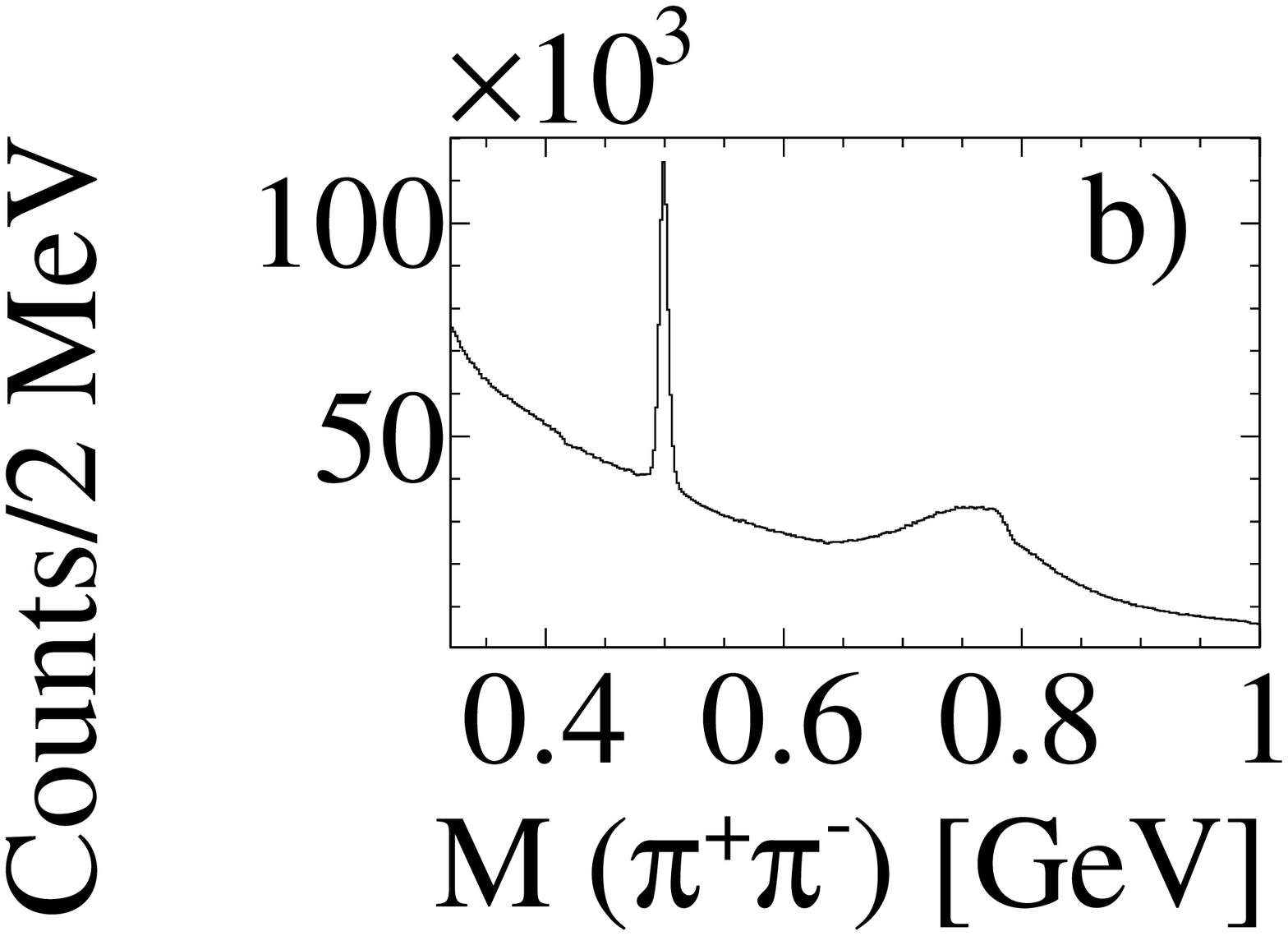}
\includegraphics[width=1.6in] {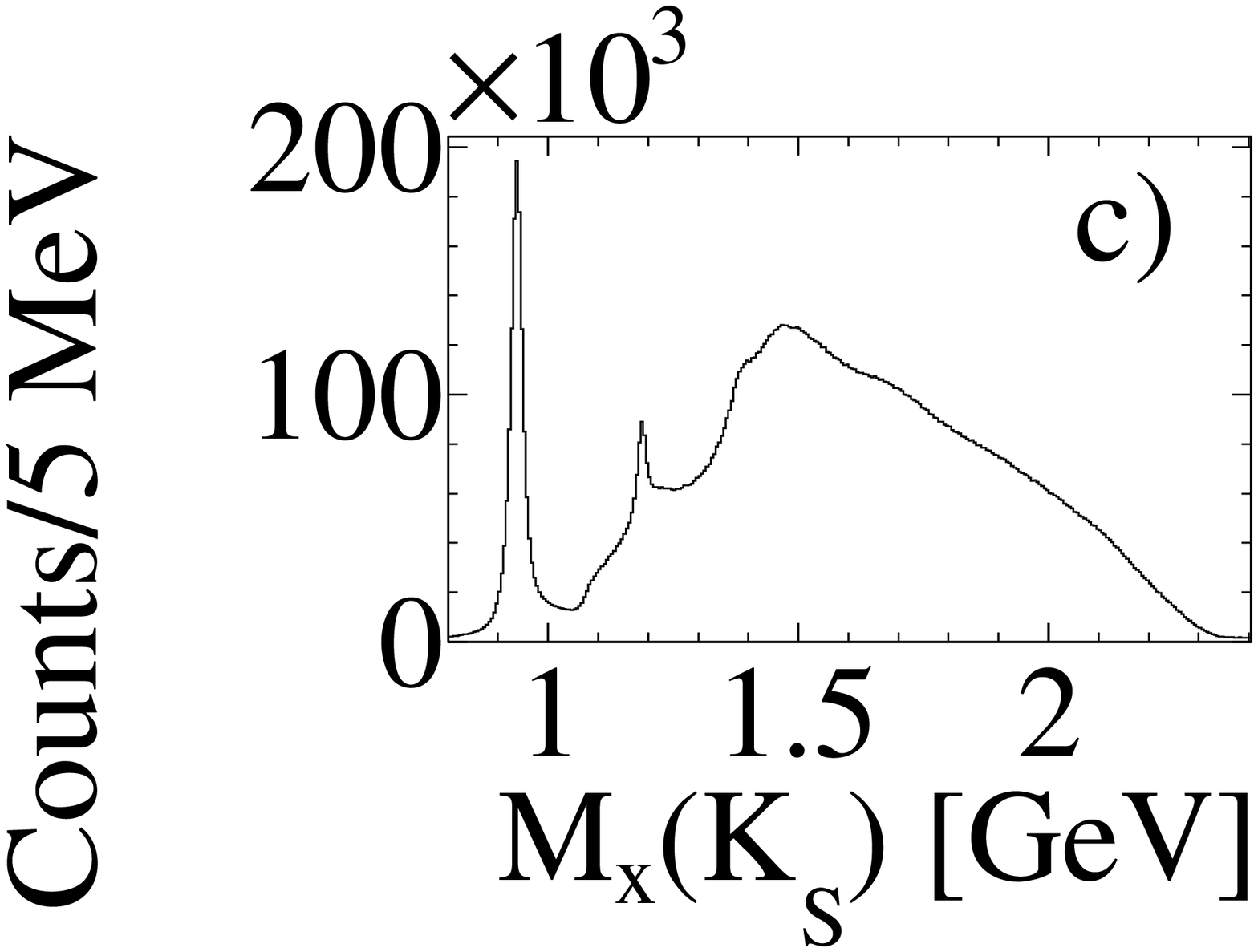}
\includegraphics[width=1.6in] {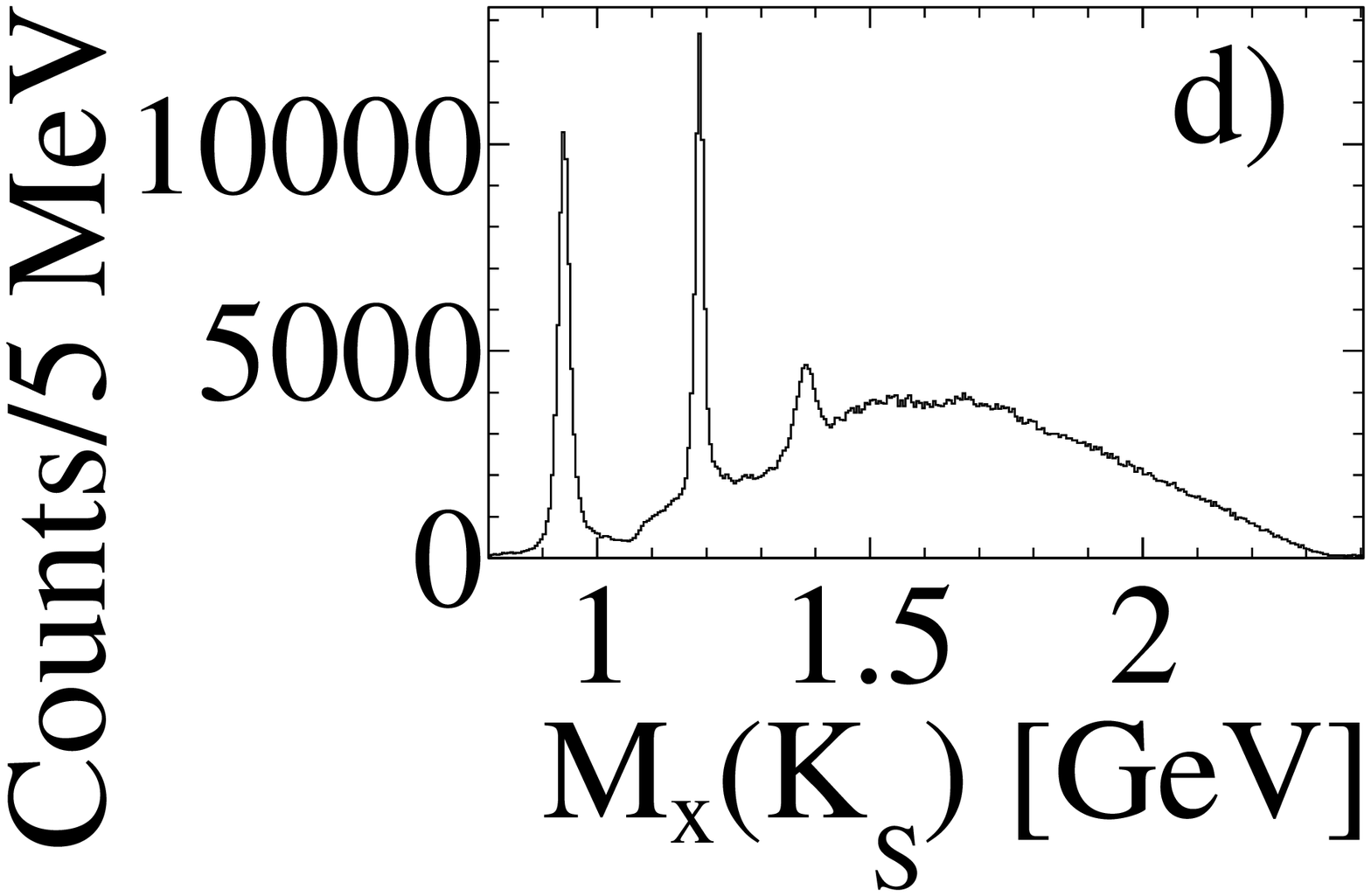}
\includegraphics[width=1.6in] {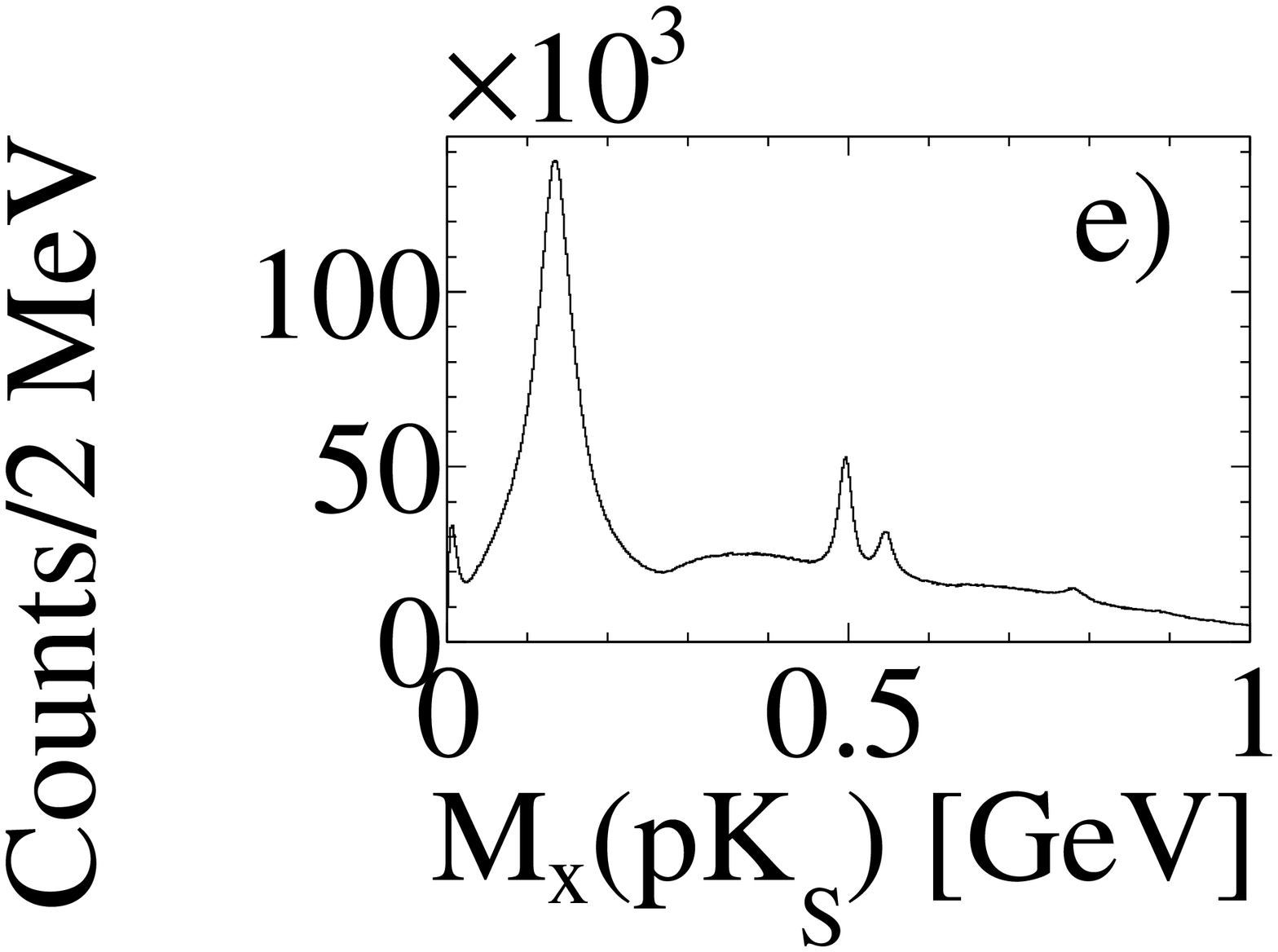}
\includegraphics[width=1.6in] {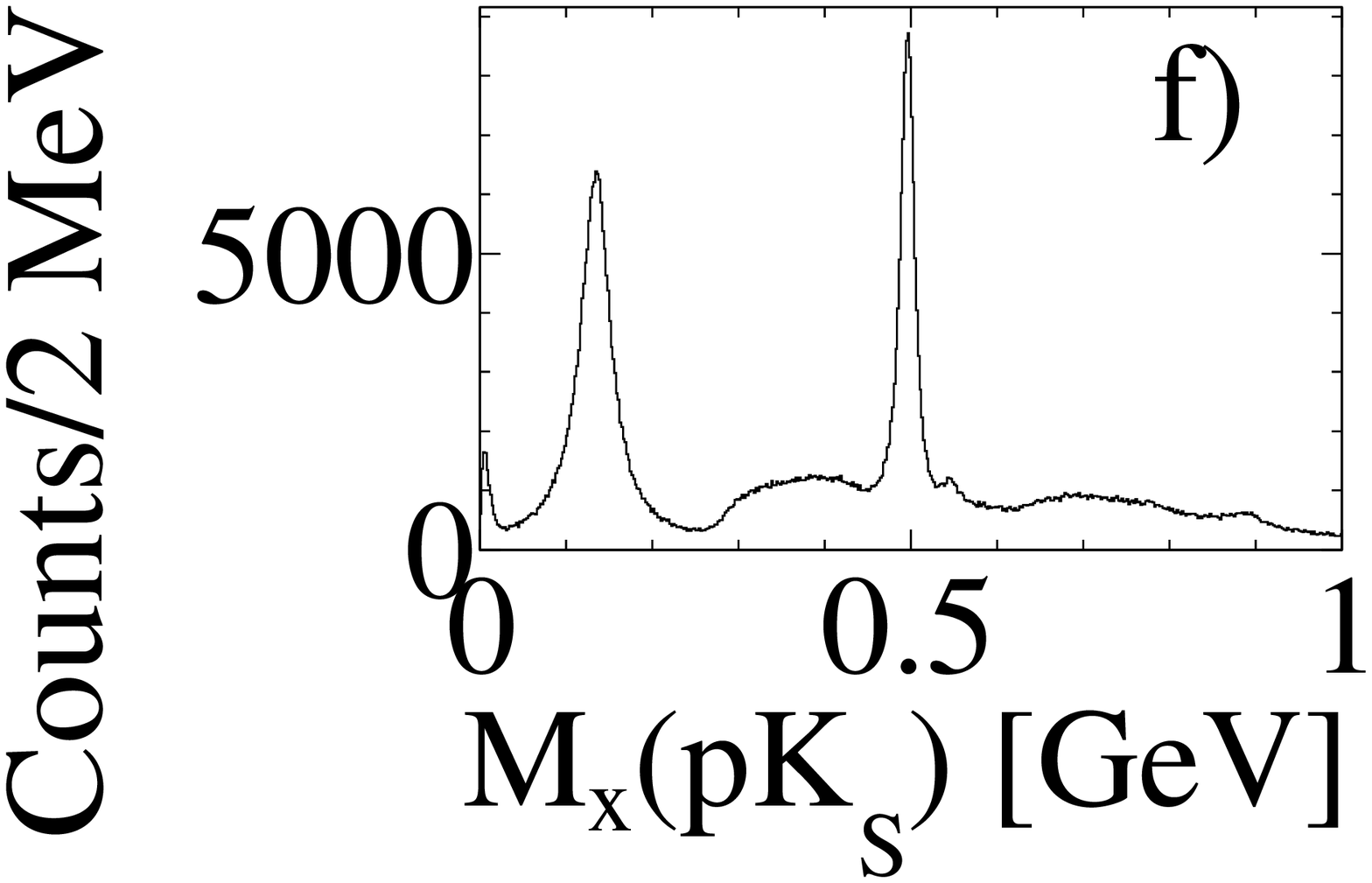}
\includegraphics[width=1.6in] {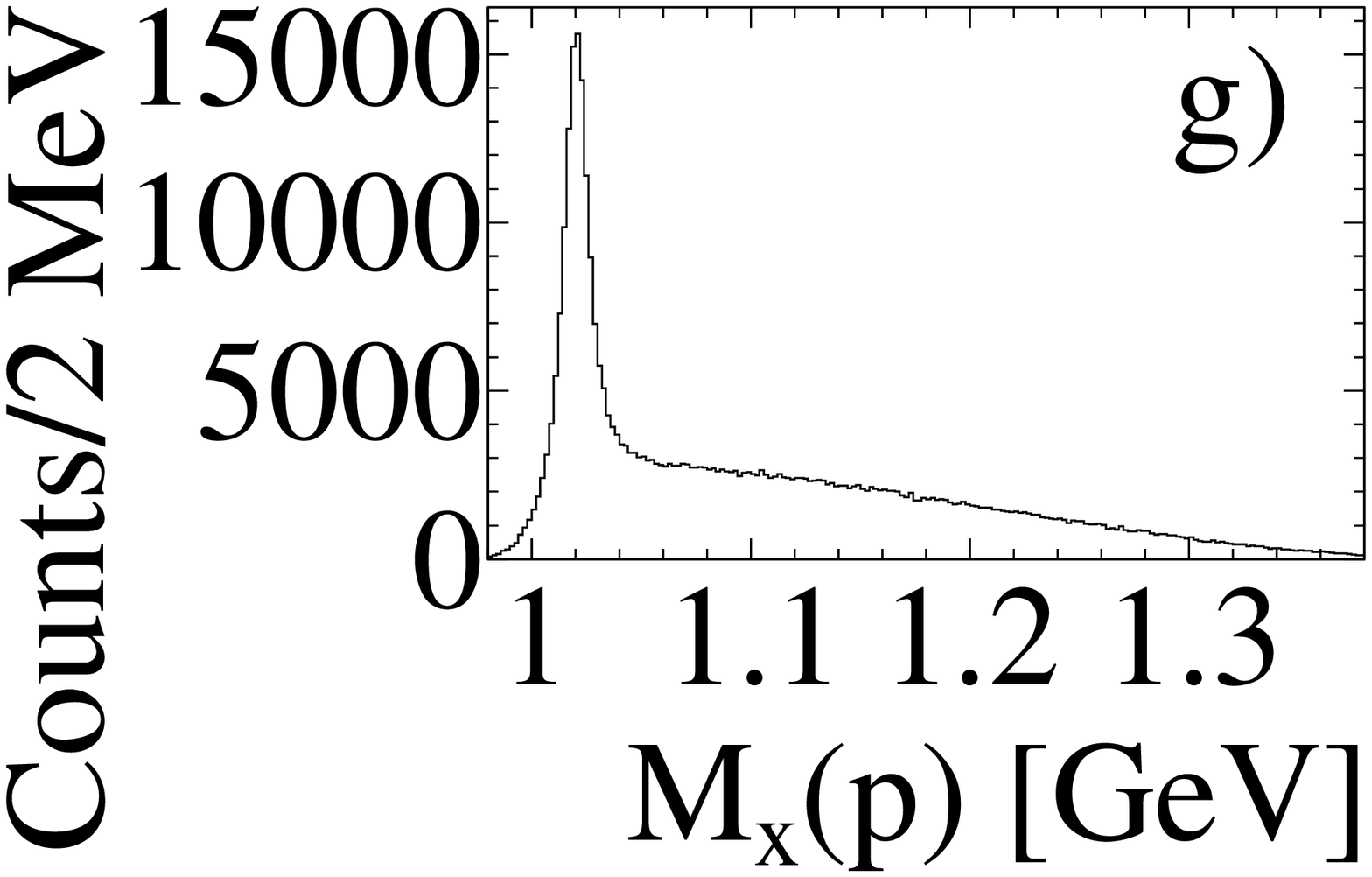}
\includegraphics[width=1.6in] {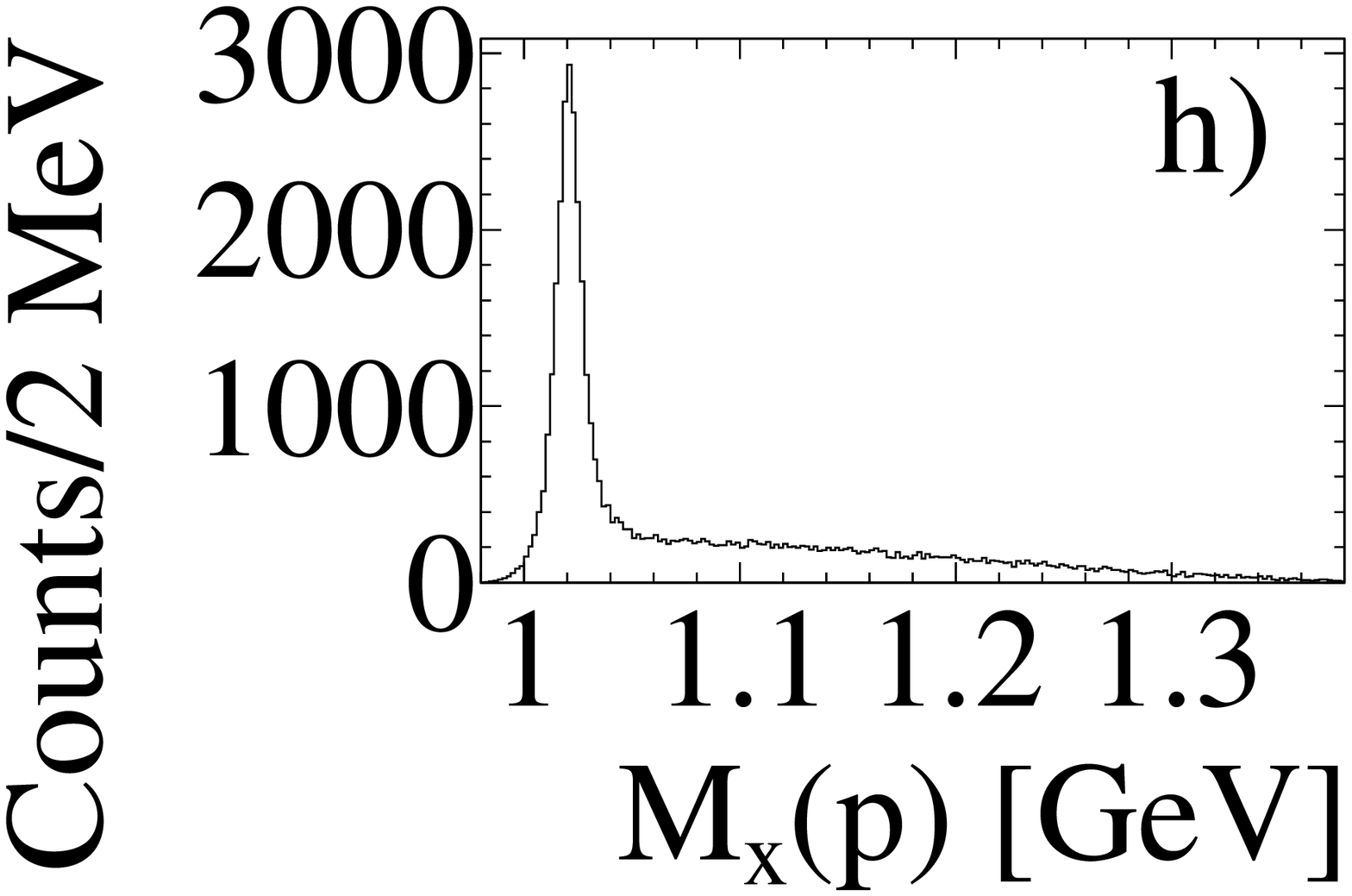}
\includegraphics[width=1.6in] {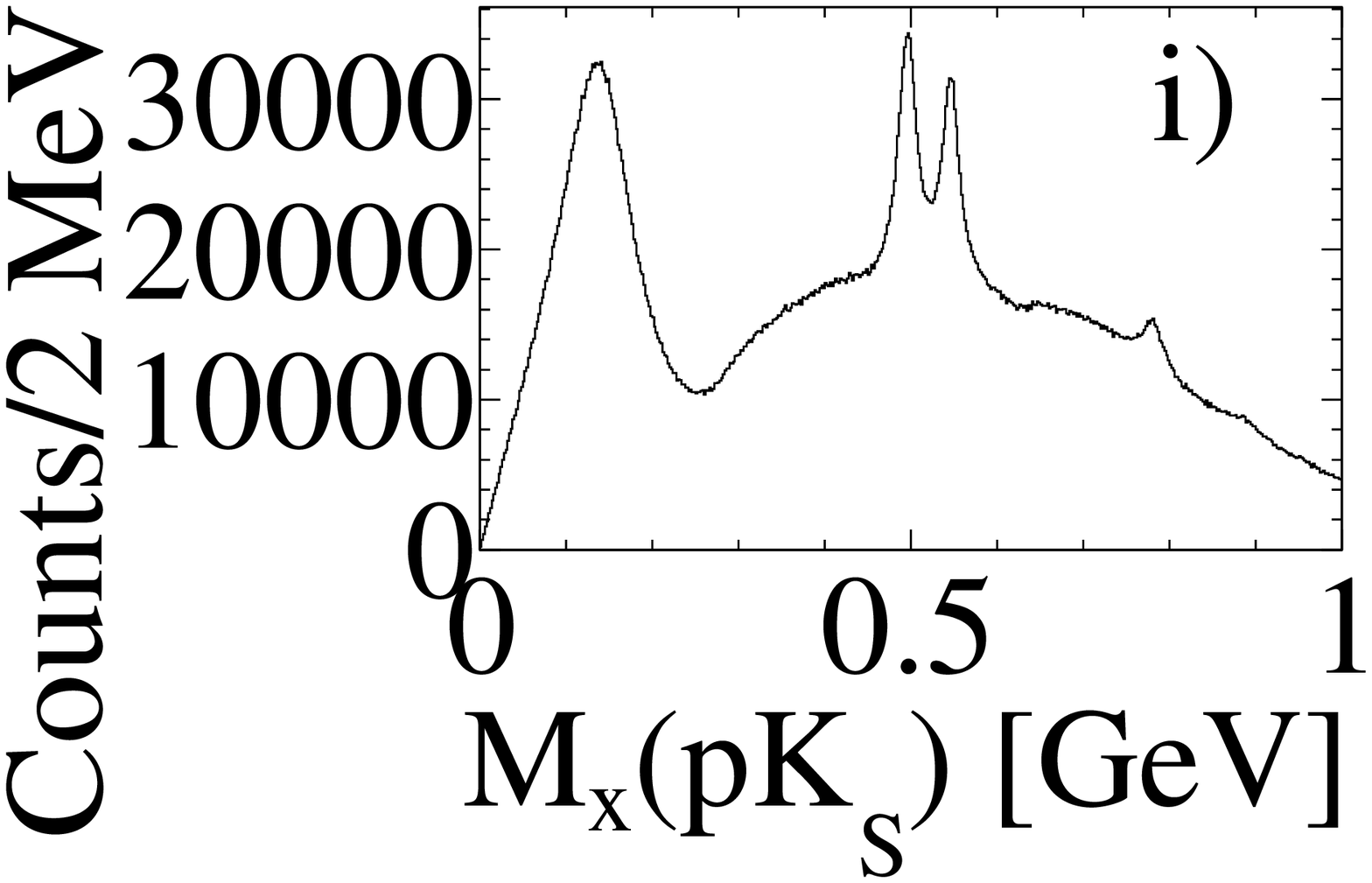}
\includegraphics[width=1.6in] {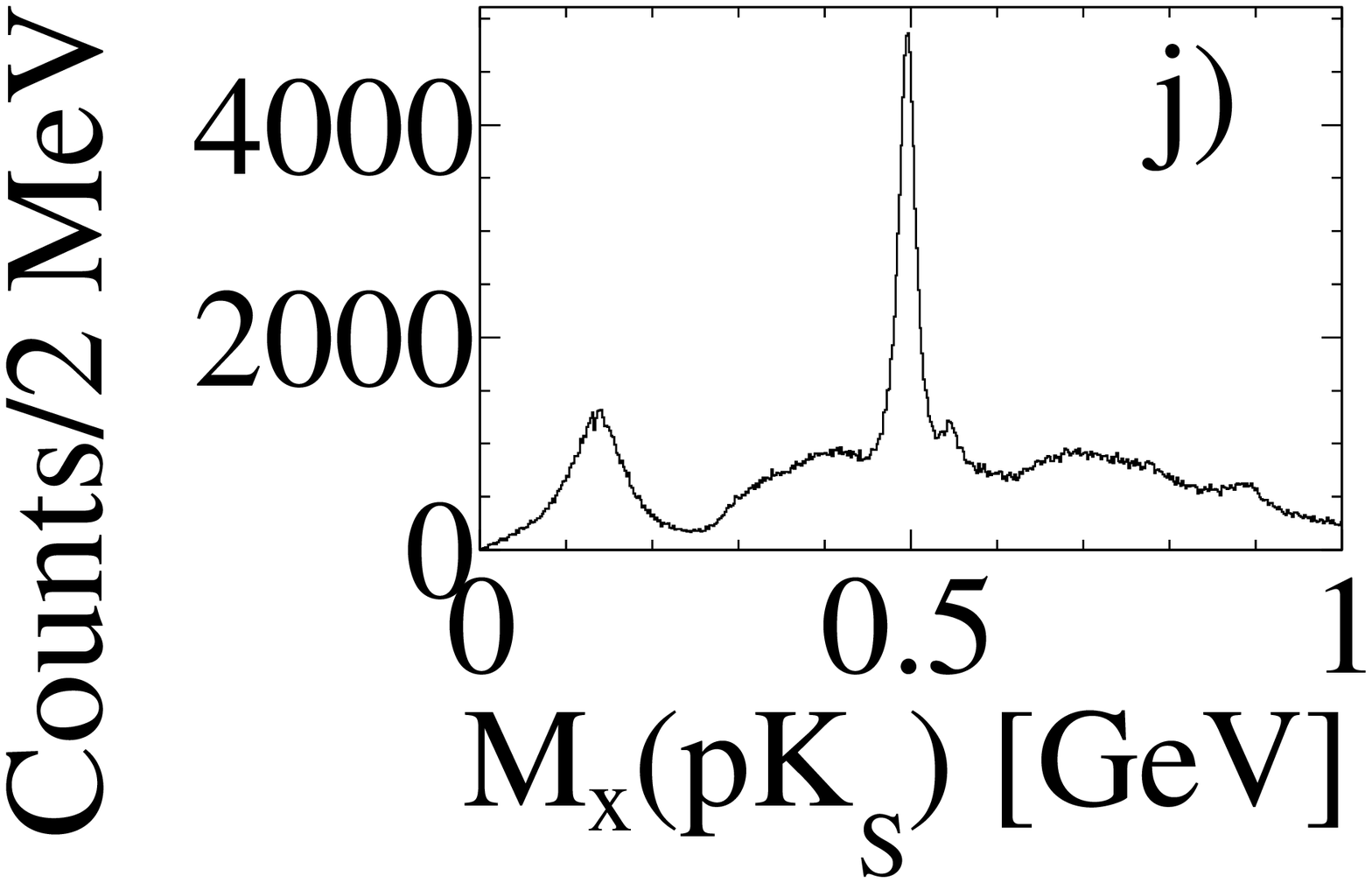}
\caption{ Upper row: invariant mass of oppositely charged pions; 
second row: missing mass of $K_S$; third row: missing mass of $pK_S$ system; 
forth row: missing mass of the proton with the cuts $M(\pi^+\pi^-)=0.497\pm$0.01~GeV and 
$M_{X}(pK_S)=0.497\pm$0.02~GeV;
fifth row: missing mass of $pK_S$ system for events above $M_{X}(p)>$1.04~GeV.
All figures in the left column are without the vertex cuts and all figures in the right column are with the vertex cuts.}
\label{fig:m_$K_S$}
\end{figure}

\begin{itemize}
\item The upper row  shows the invariant mass of the two pions with $K_S$ at $\sim0.5$ GeV. 
The collinearity cut preferentially selects events with a separated
$K_S\to \pi^+\pi^-$ decay vertex.  This reduces the $K_S$ signal by roughly a
factor of $\sim$2, but reduces the nonstrange $\pi^+\pi^-$
continuum by a factor of $\sim$30.  Thus the $\rho$ peak at 0.76 GeV
is prominent in the left plot and the $K_S$ peak at 0.5 GeV is prominent in the right-hand  plot.

\item In the second row, the missing mass of $K_S$, $M_{X}(K_S)$, $M_{X}(K_S)^2=(P_{\gamma}+P_{p}-P_{K_S})^2$, where $P_{i} (i=\gamma, p, K_S)$ is the four momentum of a given particle, is plotted by selecting
events within $M_{\pi^+\pi^-}=0.497\pm0.010$~GeV.
In the right panel, Fig. 2d, one sees prominent peaks for the proton, $\Sigma(1189)^+$ and $\Sigma(1385)^+$
states
while on the left panel (Fig.  2c, without the vertex cuts) the
$\Sigma(1385)^+$ is hardly visible.  Moreover, the vertex cuts
substantially enhance the $\Sigma(1189)^+$ signal relative to the proton peak
in Fig. 2d compared to Fig. 2c.  This is a consequence of the fact that the
vertex cuts, particularly the collinearity cut, strongly reduce the non-resonant
$\pi\pi$ continuum, which is not associated with strangeness production.

\item In the third row we show the missing mass of the proton and
  $K_S$, $M_{X}(pK_S)$, $M_{X}(pK_S)^2=(P_{\gamma}+P_{p}-P_{K_S}-P_{p^{\prime}})^2$, where $P_{p^{\prime}}$ is a four momentum of the final state proton, showing the $\pi^0$, $K^0$, and $\eta$ peaks in Fig.~\ref{fig:m_$K_S$}e.
As one can see in Fig.~\ref{fig:m_$K_S$}f,  with the vertex cuts
the signal/background
ratio  of the missing kaon is significantly improved, the $\eta$ peak almost vanishes, and
the $\pi^0$ peak, as a decay product of $\Sigma(1189)^+\rightarrow p\pi^0$, is still
prominent.
\item The fourth row shows the missing mass of the proton, $M_X(p)$,
 $M_X(p)^2=(P_{\gamma}+P_{p}-P_{p^{\prime}})^2$,  by selecting
$K_S$ ($M_{\pi^+\pi^-}=0.497\pm0.010$~GeV) and $K_L$
($M_{X}(pK_S)= 0.497\pm0.020$~GeV) from the first and third rows.
One can see a peak for the
$\phi$ meson in both cases: without (Fig.~\ref{fig:m_$K_S$}g), and with (Fig.~\ref{fig:m_$K_S$}h),
vertex cuts. Again  the signal/background ratio is significantly
improved with the
vertex cuts.
\item Finally, in the fifth row (Fig.~\ref{fig:m_$K_S$}i and Fig.~\ref{fig:m_$K_S$}j) we show again the missing mass of the proton
and $K_S$, $M_X(pK_S)$, this time plotted only for events outside the $\phi$-peak,
{\em i.e.} $M_X(p)>1.04$.  The left panel corresponds to the event selection used in the previous CLAS analysis of these data~\cite{CLAS1}. 

\end{itemize}

From Fig.~\ref{fig:m_$K_S$} one can conclude that the application of the
vertex cuts significantly improves the identification of the
final state particles.

\begin{figure}[htb!]
\includegraphics[width=3.2in] {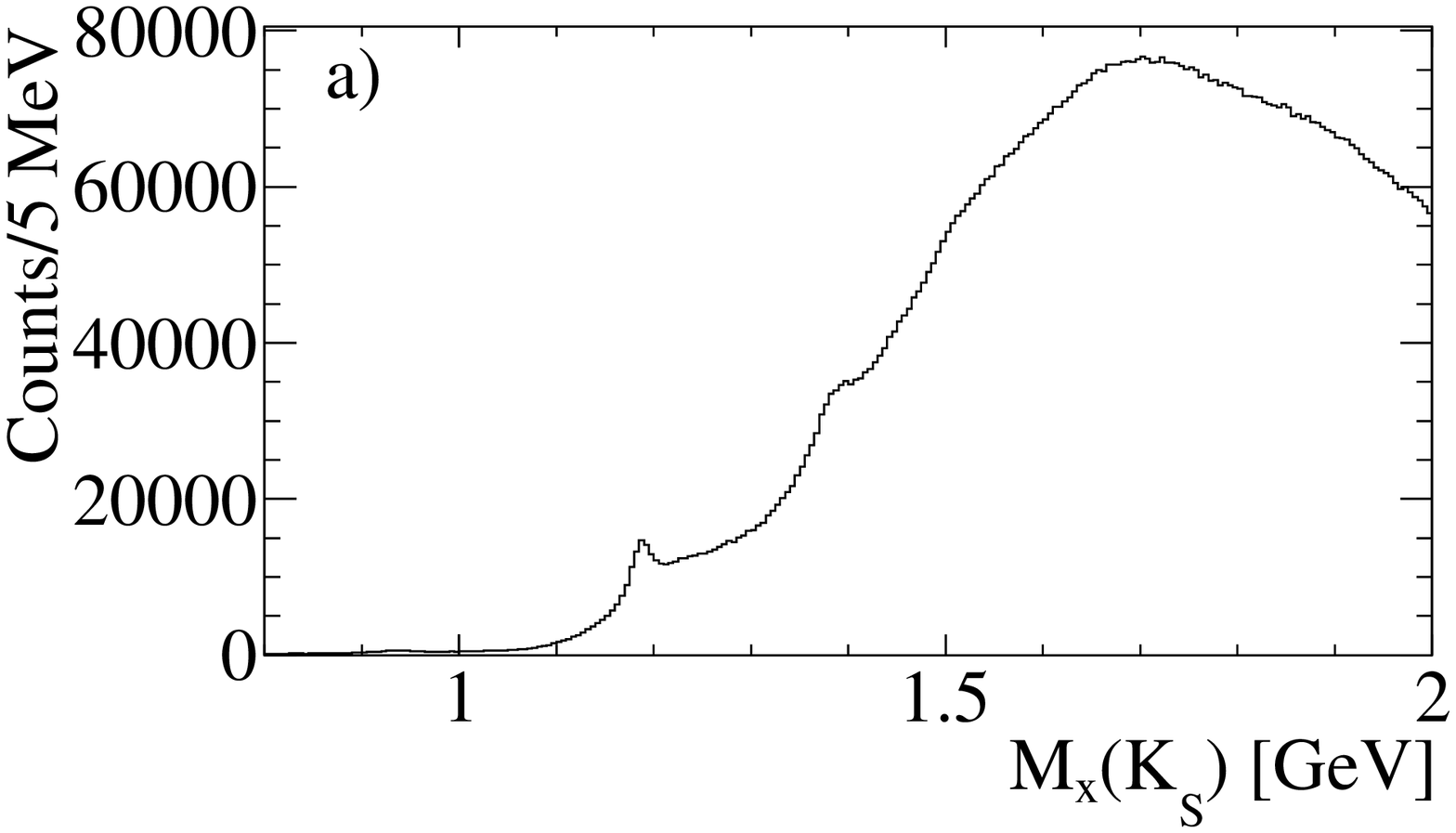}
\includegraphics[width=3.2in] {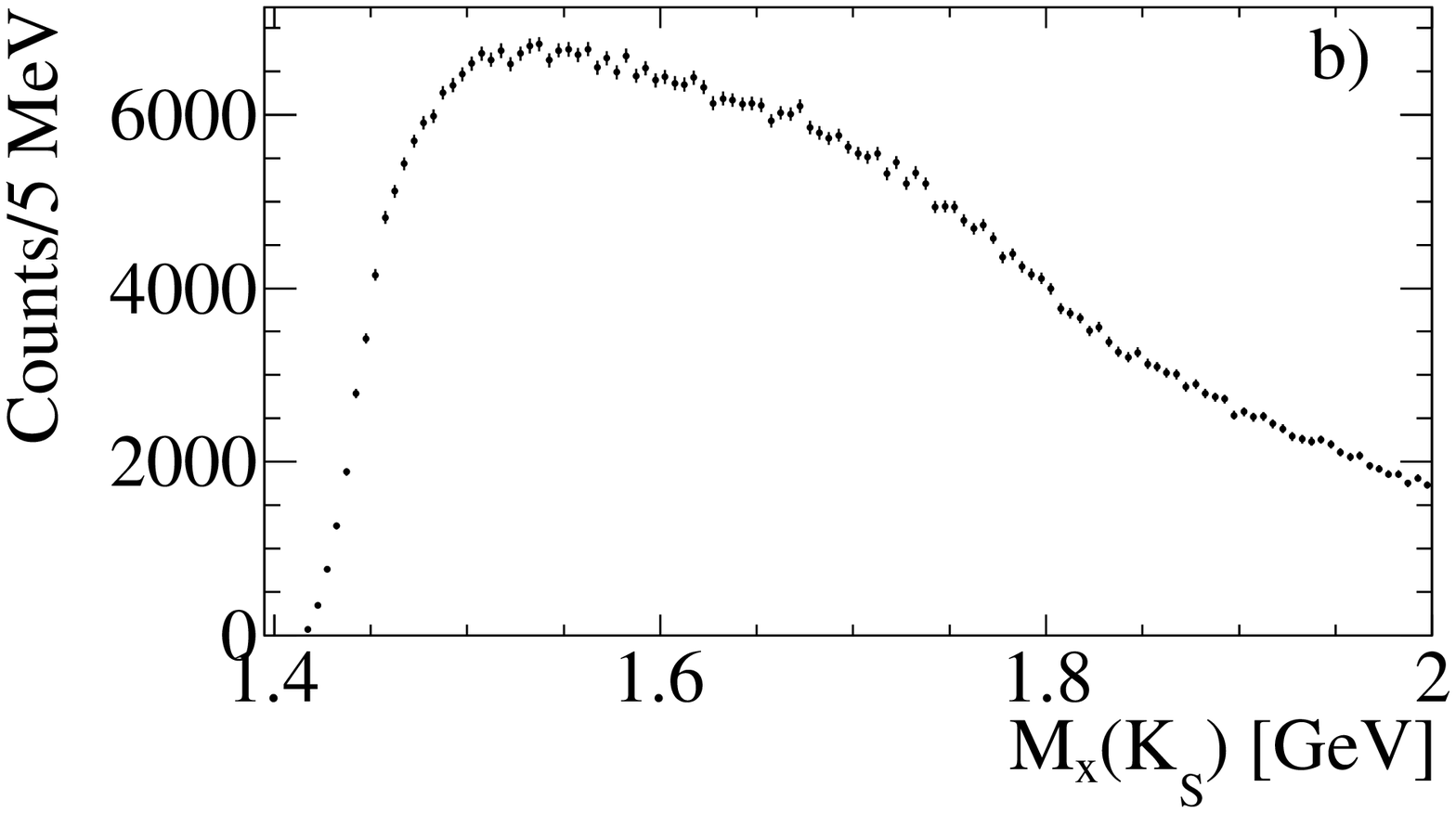}
\caption{Upper panel: missing mass of $K_S$ without vertex cuts
  and no cut  on the $K_L$ peak. Lower panel: missing mass of $K_S$
  without vertex cuts, but with the cut  on the $K_L$ peak. 
Both histograms are for events selected above the $\phi$ peak $M_{\phi}>$1.04~GeV.
}
\label{fig:mm_ks_old}
\end{figure}

In Fig.~\ref{fig:mm_ks_old} the missing mass of $K_S$ is presented without
vertex cuts and for events above the $\phi$ peak, $M_{\phi}>$1.04~GeV.
The upper panel, Fig.~\ref{fig:mm_ks_old}a, is for events without a cut
on the $K_L$ peak. 
Although there are many events
in the distribution, a
prominent state such as the $\Sigma(1385)^+$ is barely visible on top
of a very high
background. By applying an additional cut on the $K_L$ peak we
reproduce the CLAS published  analysis~\cite{CLAS1}, and obtain
a similar structureless distribution, as presented in
Fig.~\ref{fig:mm_ks_old}b. The upper limit of the $\Theta^+$
photoproduction cross section in~\cite{CLAS1}
 was estimated from this distribution.

\subsection{Interference with $\phi$ production}

In this section we present our results for $pK_LK_S$ events selected under the $\phi$ peak.
The main goal is to study the missing mass of $K_S$, which is equivalent to the invariant mass of the final state proton and missing $K_L$, $M_X(K_S)=M(pK_L)$.
This kinematic  domain has not been studied before and,  as 
discussed in the introduction, might possibly reveal a tiny signal in
the missing mass of $K_S$ due to interference with the very strong
signal of $\phi$ production. 

\begin{figure}[htb!]
\includegraphics[width=3.2in]{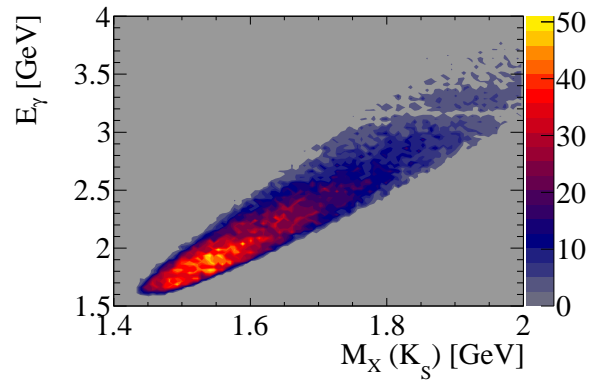}
\caption{(Color online). Incoming photon beam energy versus $M_{X}(K_S)$ for events selected under the $\phi$ peak.}
\label{fig:eg_mpkl}
\end{figure}

In Fig.~\ref{fig:eg_mpkl} the incoming photon beam energy is plotted
versus $M_{X}(K_S)$ for events selected under the $\phi$ peak with a cut $M_X(p)=1.02\pm0.01$~GeV.
Photons above 2.4~GeV do not contribute to the region of
$M_X(K_S)$=(1.5-1.6)~GeV, where previously signals for a resonance in the
$KN$ system have been reported. Therefore in the following we used
data with a safe upper limit cut $E_{\gamma}<2.6$~GeV, which at the
same time includes a sufficiently wide
range of photon energy to control  the phase space distribution of $\phi$ production.

\begin{figure}[htb!]
\includegraphics[width=3.0in] {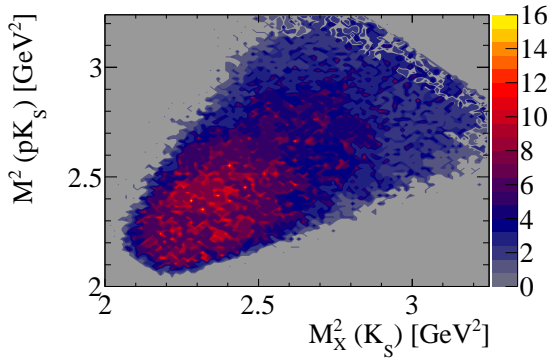}
\caption{(Color online). Dalitz plot:~invariant mass $M^2(pK_S)$ versus $M^2_{X}(K_S)$ for events selected under the $\phi$ peak.}
\label{fig:mpks_mpkl}
\end{figure}

In order to see the whole kinematic phase space, in
Fig.~\ref{fig:mpks_mpkl} we present the  Dalitz plot, $M(pK_S)^2$
versus $M(pK_L)^2$, for events selected under the $\phi$ peak,
$M_X(p)=1.02\pm0.01$~GeV. To perform a search for a resonance
structure in the missing mass of $K_S$, i.e. $M(pK_L)$, we need to
restrict kinematic overlap with another system created in the
invariant mass $M(pK_S)$, such as the well known $\Sigma^*$ resonances,
which could affect and wash out a possible signal for a narrow structure.

\begin{figure}[htb!] 
\includegraphics[width=3.4in]{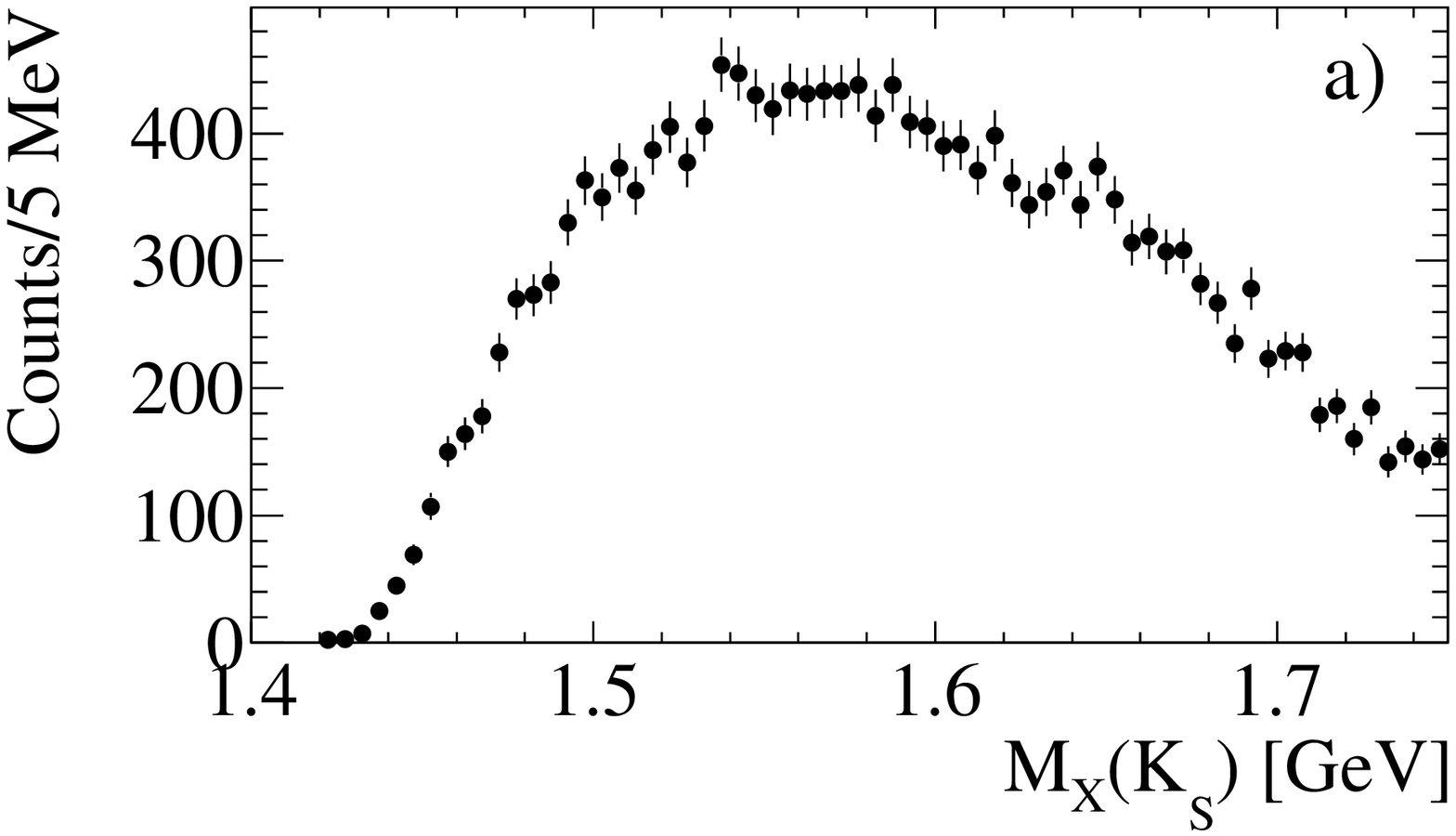}
\includegraphics[width=3.4in]{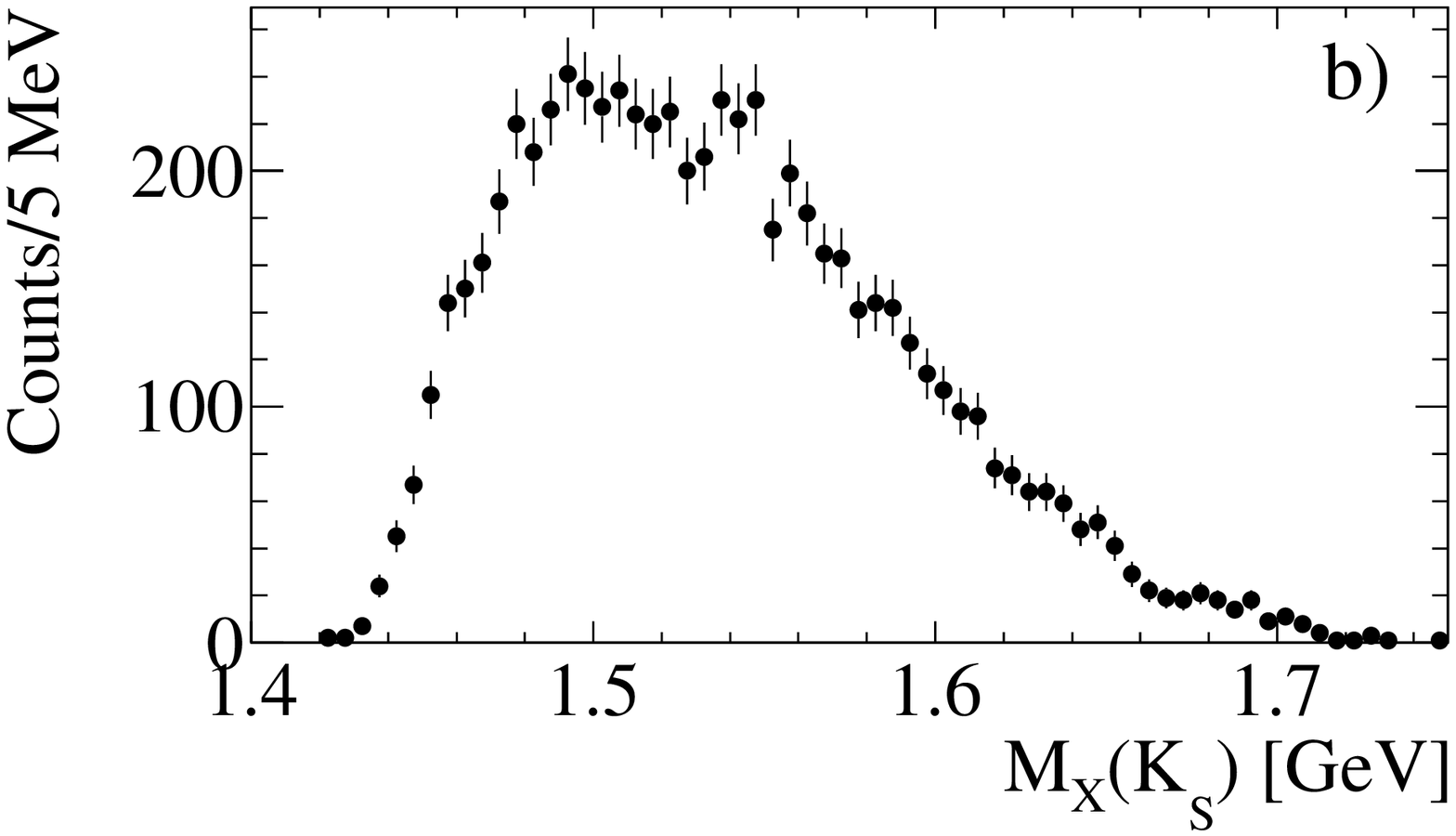}
\includegraphics[width=3.4in]{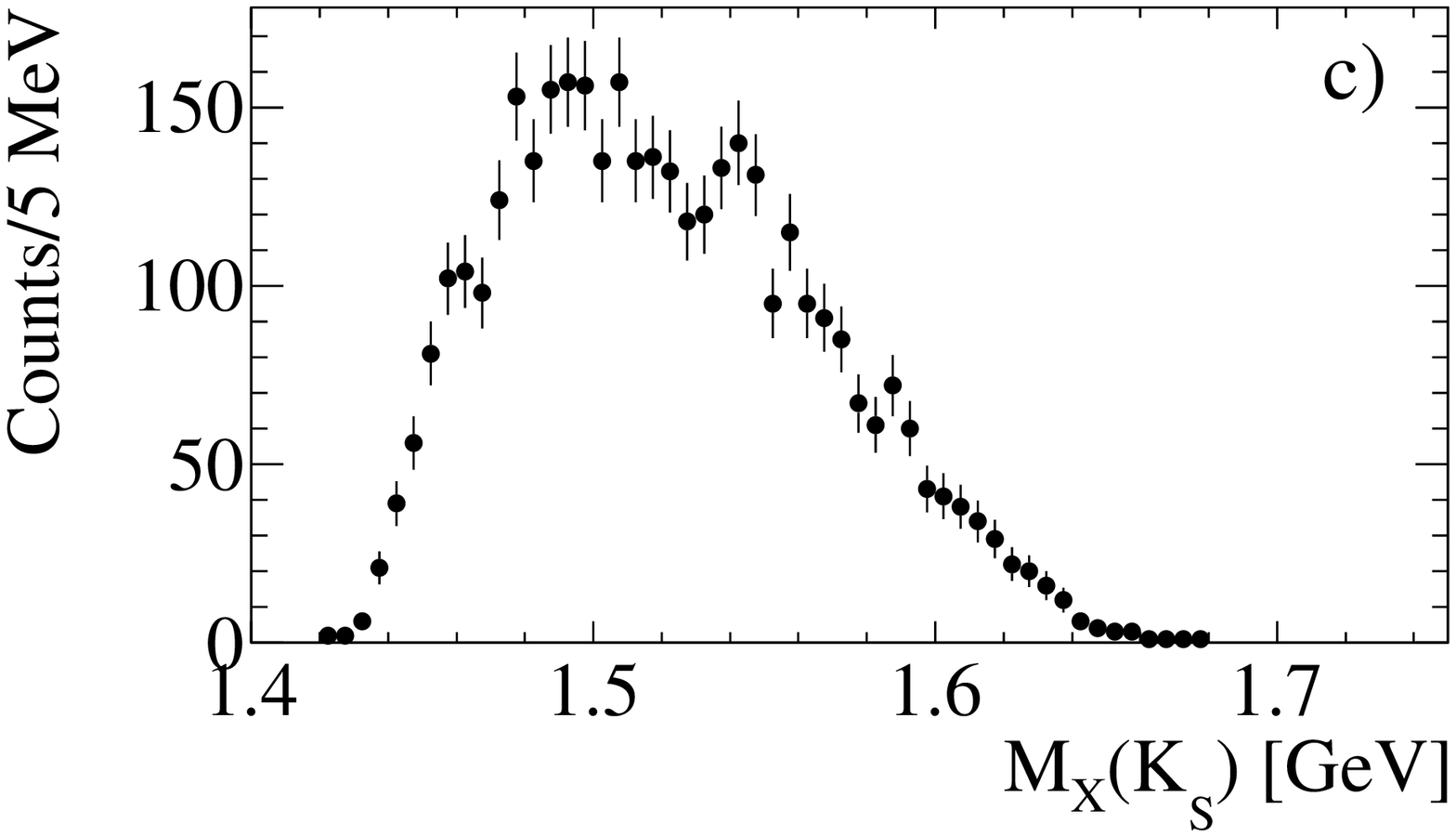}
\includegraphics[width=3.4in]{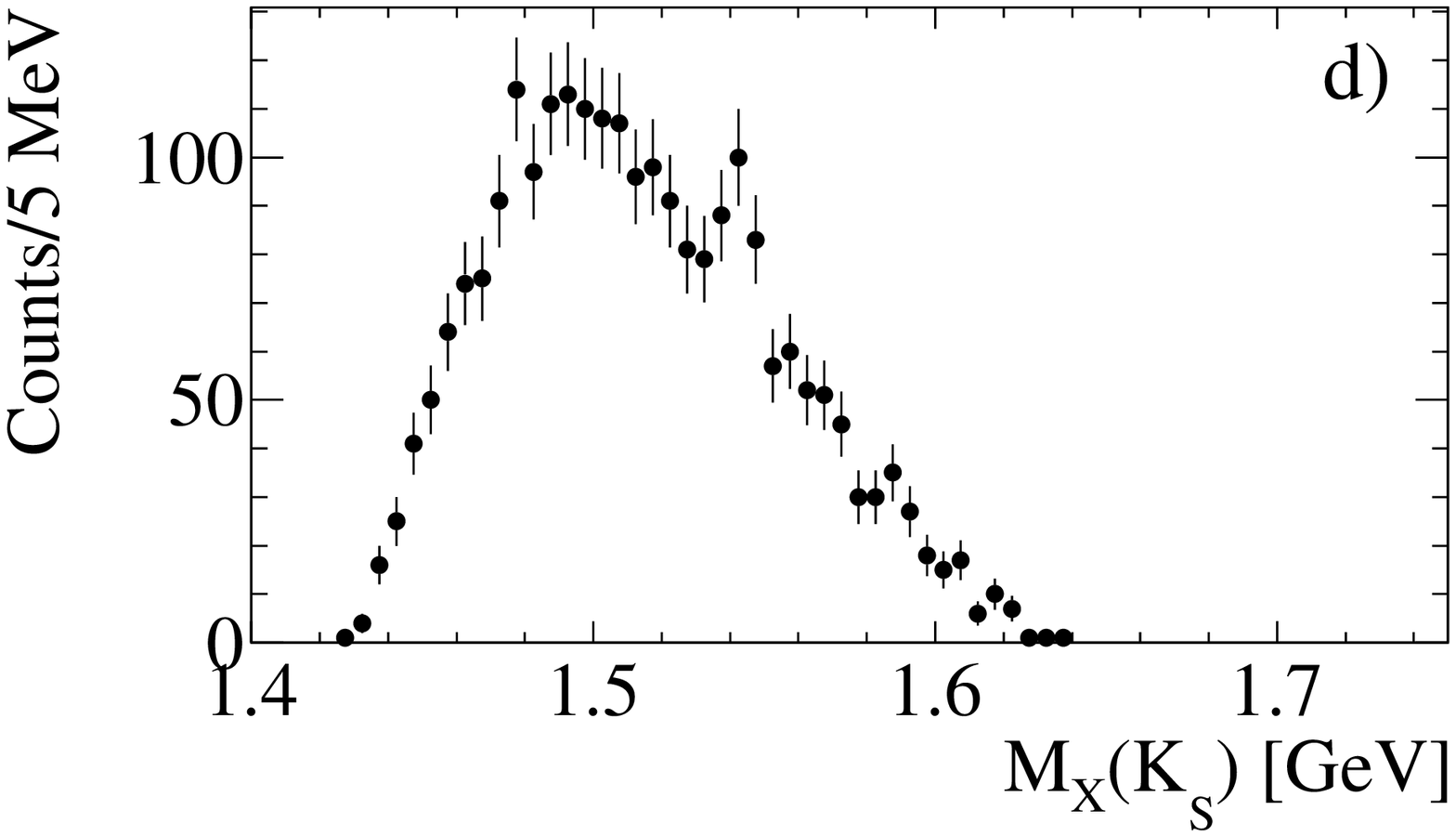}
\caption{Missing mass of $K_S$ is plotted with different cuts on the invariant mass $M(pK_S)$.
a)  no cuts (vertex cuts only), b) $M(pK_S)<1.56$~GeV, c)  $M(pK_S)<1.52$~GeV, 
and d) $M(pK_S)<1.5$~GeV.}
\label{fig:mpkl_cut_mpks}
\end{figure}

In Fig.~\ref{fig:mpkl_cut_mpks} $M_X(K_S)$ is presented with different
cuts on the invariant mass $M(pK_S)$, namely 
no cut (vertex cuts only)  Fig.~\ref{fig:mpkl_cut_mpks}a with total
number ov events $N_{events}$=20007, with the cut
$M(pK_S)<1.56$~GeV in Fig.~\ref{fig:mpkl_cut_mpks}b with
$N_{events}$=6766 , with the cut $M(pK_S)<1.52$~GeV in
Fig.~\ref{fig:mpkl_cut_mpks}c with $N_{events}$=3744,
and with the cut $M(pK_S)<1.5$~GeV in Fig.~\ref{fig:mpkl_cut_mpks}d
with $N_{events}$=2380.
As one can see  there are hints of some structure around 1.54~GeV. By gradually changing the cut on $M(pK_S)$ the peak structure in 
$M_X(K_S)$ becomes more and more prominent.

\begin{figure}[htb!]
\includegraphics[width=3.0in] {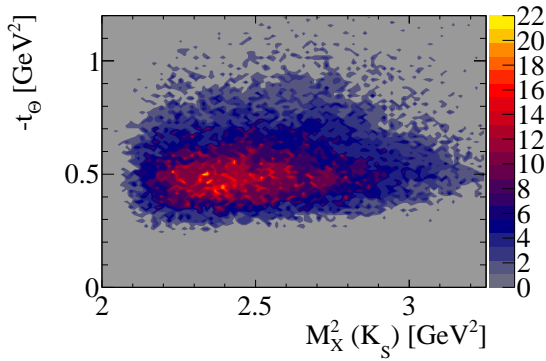}
\caption{(Color online). Chew-Low diagram:~$t_{\Theta}$ versus $M^2_{X}(K_S)$ for events selected under the $\phi$ peak.}
\label{fig:Chew-Low}
\end{figure}

\begin{figure}[htb!]
\includegraphics[width=3.4in]{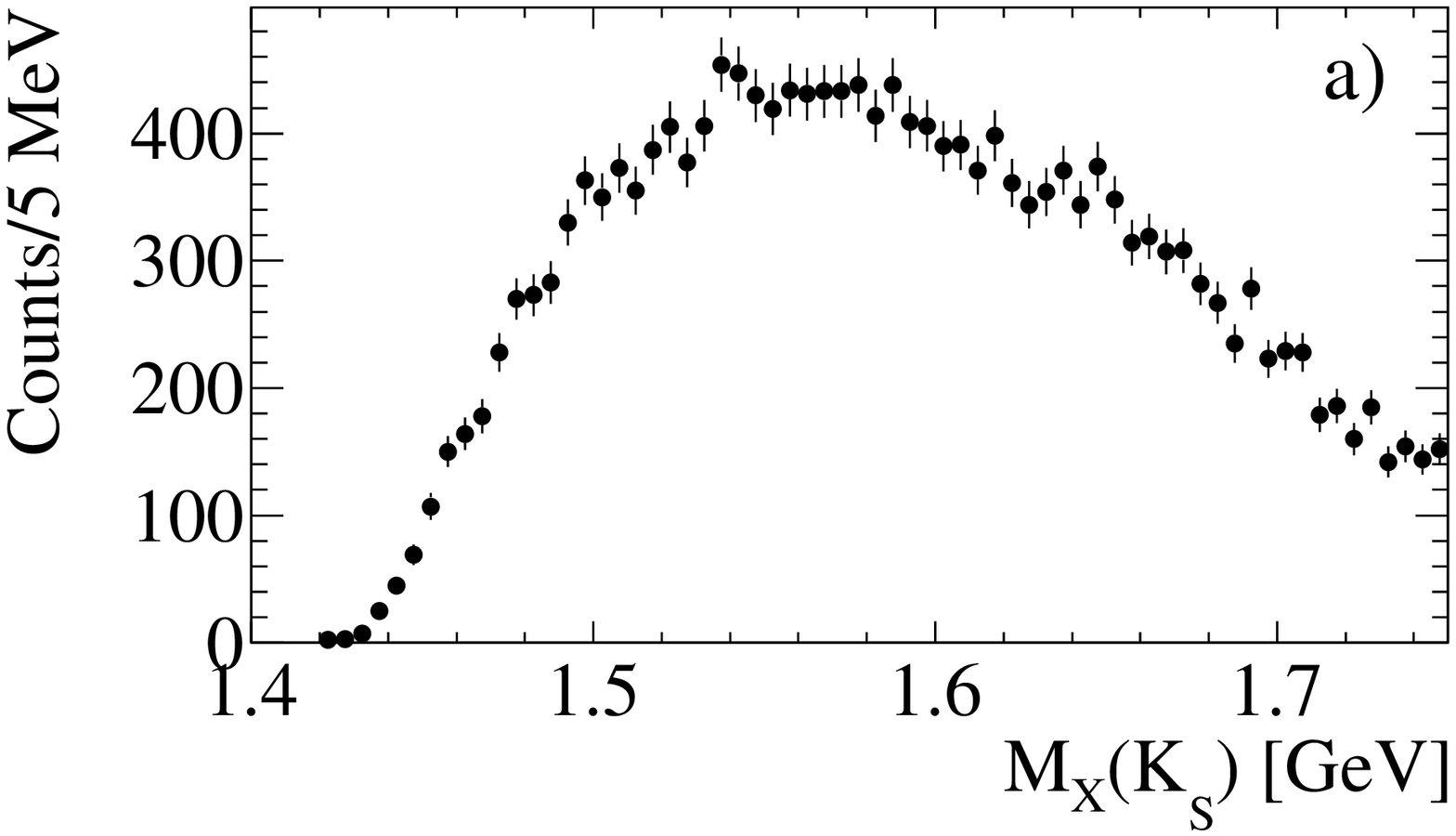}
\includegraphics[width=3.4in]{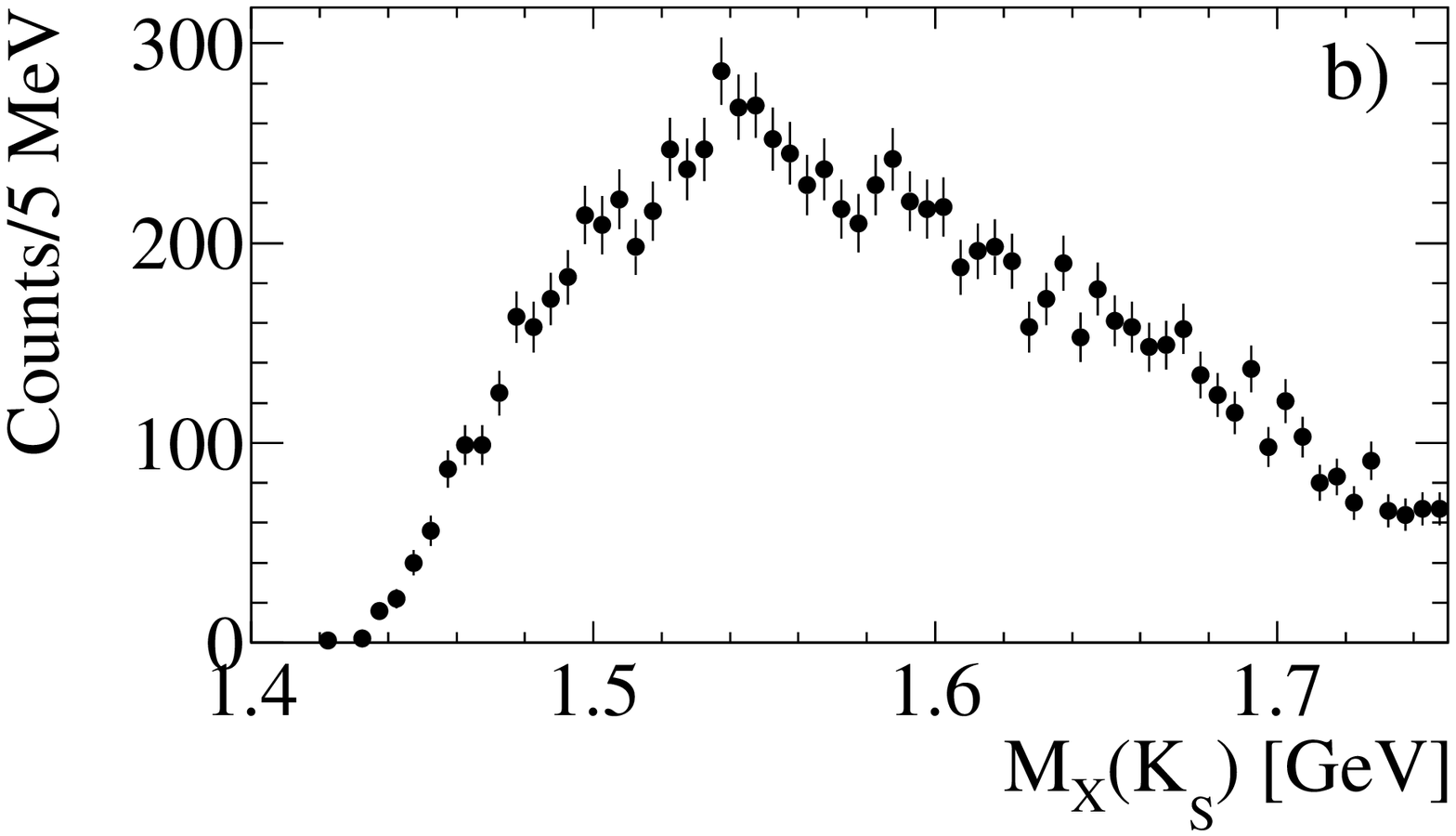}
\includegraphics[width=3.4in]{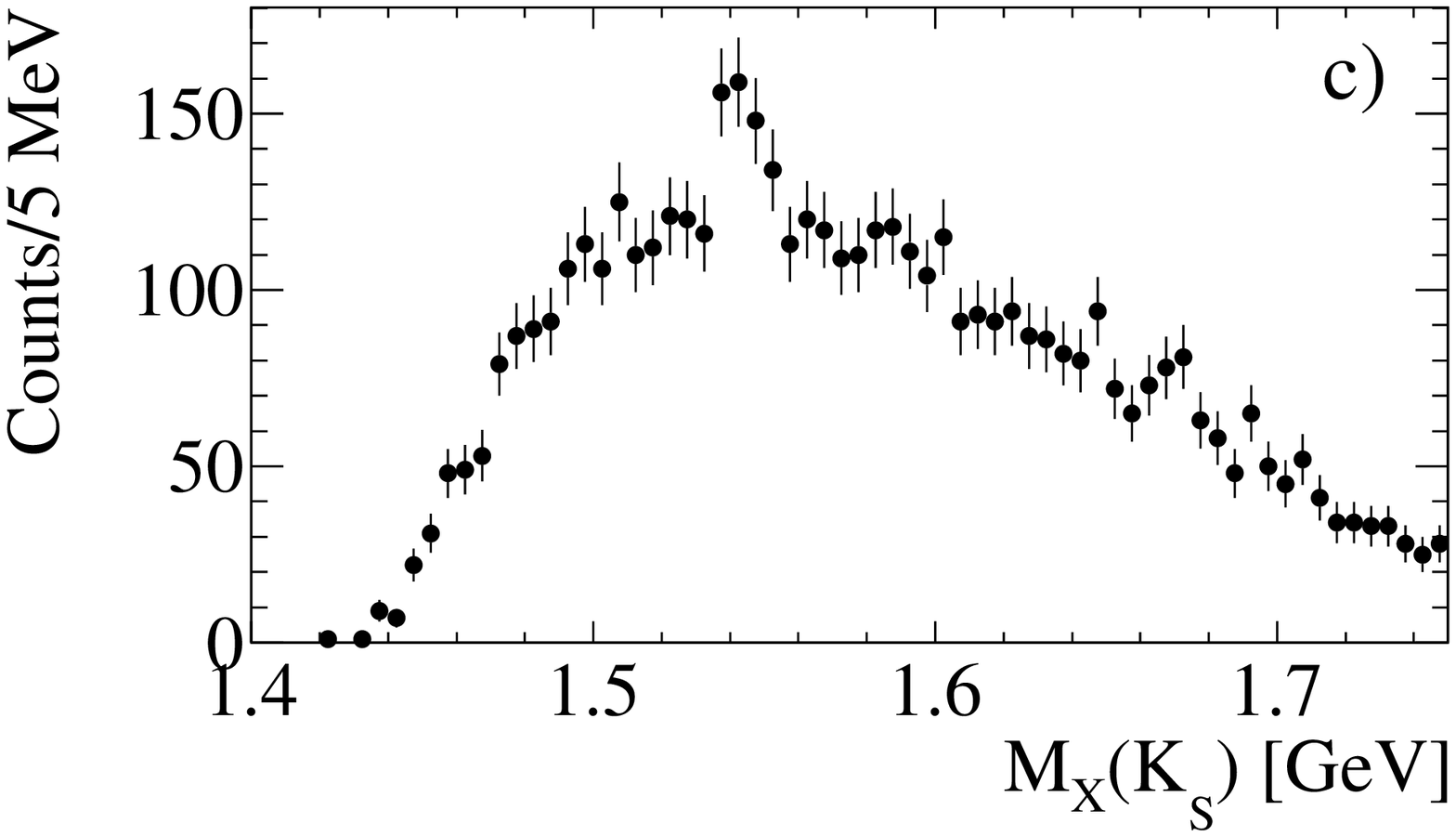}
\includegraphics[width=3.4in]{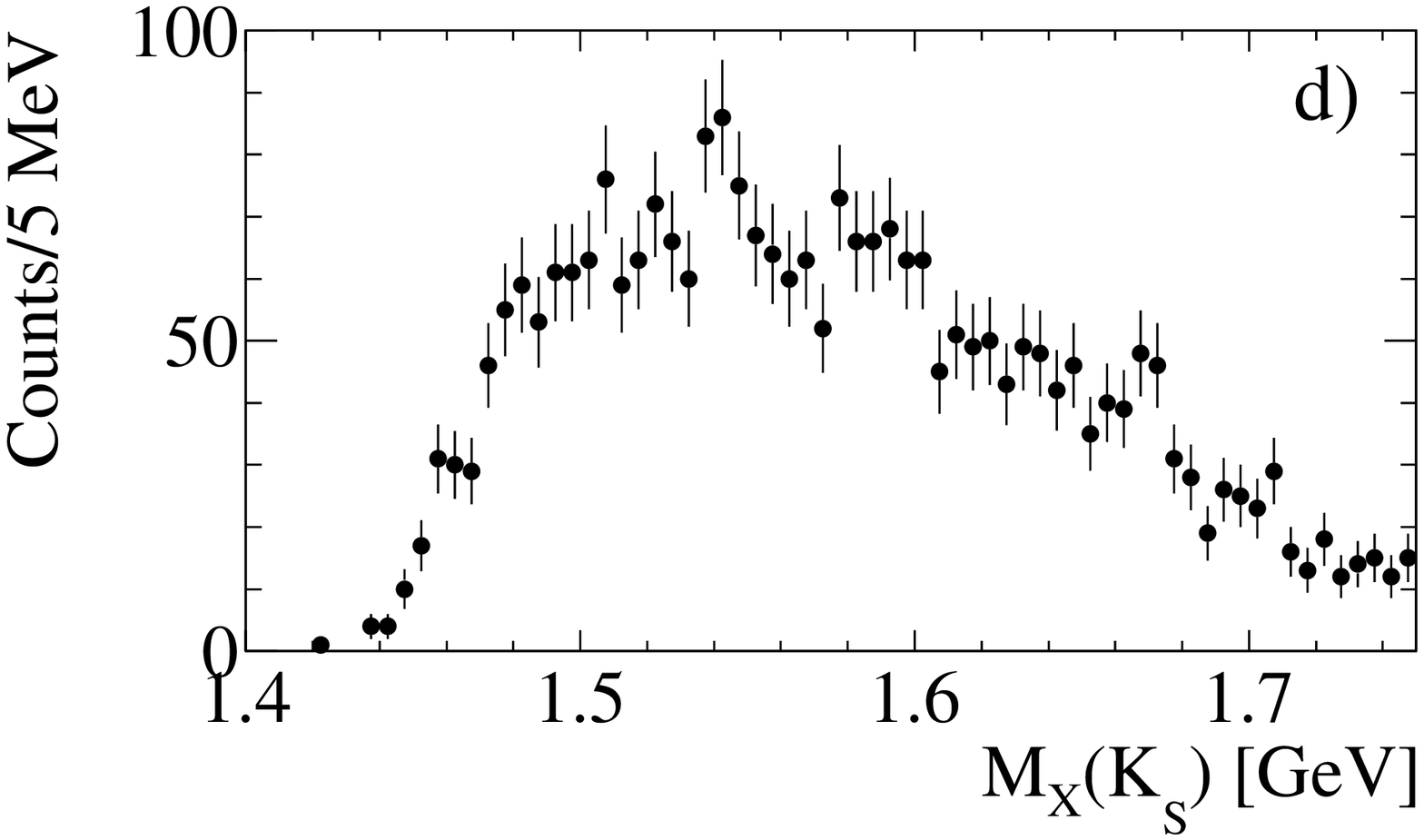}
\caption{Missing mass of $K_S$ with different cuts on $t_{\Theta}$: 
a) with no cut on $t_{\Theta}$, b) $-t_{\Theta}<0.55$~GeV$^2$ , 
c) $-t_{\Theta}<0.45$~GeV$^2$, and  d) $-t_{\Theta}<0.4$~GeV$^2$.}
\label{fig:mpkl_t_cut}
\end{figure}

Next in Fig.~\ref{fig:Chew-Low} we plot the Chew-Low diagram, $t_{\Theta}$ versus $M_{X}(K_S)$.
Here $t_{\Theta}$ is defined as $t_{\Theta}= (P_{\gamma}-P_{K_S})^2$, where $P_{\gamma}$ and $P_{K_S}$ are four momenta of the incoming
photon and reconstructed $K_S$. Since we do not know the mechanism for photoproduction of the possible
resonance in the $M(pK_L)$ system, we assumed that it should be produced with some exponential $t$-dependence, like other baryons,
such as $\Lambda(1520)$. Therefore we expected that by selecting lower $t_{\Theta}$-values we would suppress 
the background without losing too many signal events.~In Fig.~\ref{fig:mpkl_t_cut}  the distribution of
$M_{X}(K_S)$ is presented without a cut on $t_{\Theta}$ in
Fig.~\ref{fig:mpkl_t_cut}a with $N_{events}$=20007, with a cut $-t_{\Theta}<0.55$~GeV$^2$  in
Fig.~\ref{fig:mpkl_t_cut}b with $N_{events}$=10590, with a cut
$-t_{\Theta}<0.45$~GeV$^2$ in Fig.~\ref{fig:mpkl_t_cut}c with $N_{events}$=5271, and with a cut $-t_{\Theta}<0.4$~GeV$^2$
in Fig.~\ref{fig:mpkl_t_cut}d with $N_{events}$=2848.~Fig.~\ref{fig:mpkl_t_cut} does not include a $M(pK_S)$ cut. The statistical significance of the structure at $\sim$1.54~GeV is
maximized for values of $-t_{\Theta}<0.45$~GeV$^2$. By applying a
tighter cut, $-t_{\Theta}<0.4$~GeV$^2$ (Fig.~\ref{fig:mpkl_t_cut}d), we lose 
statistics and the statistical significance of the observed structure deteriorates.~We note that the significance of the
structure at 1.54 GeV does not vary as one would expect purely from
the statistics.  This could be the result of a complicated
interference between the $\phi$ and the baryon resonance.
For example the $\phi$ production mechanism changes at about
$t_\phi=(P_{\gamma}-P_{\phi})^2\approx -0.5$~GeV$^2$ from predominantly diffractive (at lower $|t_\phi|$) to predominantly
$s$-channel~\cite{titov,williams}. If the phase of the interference depends on
the $\phi$ production mechanism, then integrating over different mechanisms could
wash out a possible  signal in the $pK_L$ system. Moreover, $s$-channel
$\phi$-production decreases much more slowly with $t_\phi$ than diffractive $\phi$-production.
Thus, inclusion of the $s$-channel
mechanism could further decrease the signal/background ratio.   The
cut on $t_{\Theta}$ also significantly
reduces the range of $t_\phi$  thereby
reducing the $s$-channel contribution and potentially improving the
signal/background ratio.

From Figs.~\ref{fig:mpkl_cut_mpks} and ~\ref{fig:mpkl_t_cut} one
can see that the resonance structure around 1.54~GeV appears either by
restricting the $M(pK_S)$ invariant mass or by selecting the low $t_{\Theta}$
region.

We are unable to find any significant peak in the invariant mass
spectrum $M(pK_S)$.
This is because the resolution of low
momentum protons is significantly worse in CLAS than the photon energy resolution.  The $p\,K_L$
mass is computed from the missing mass $M_X(\gamma p\rightarrow\pi^+\pi^-X)$ and depends
only on the pion and photon resolutions.  The $p\,K_S$ mass is computed from the
$p\,\pi^+\pi^-$ mass and depends on both the pion and proton resolutions. Detailed Monte
Carlo studies have shown that the CLAS resolution for the invariant mass $M(p\,K_S)$ is
much worse than for the missing mass $M_X(\gamma p\rightarrow\pi^+\pi^-X)$ due to the use
of low momenta protons in the reconstruction of the invariant mass. 
Similarly, a generated narrow peak is not reconstructed as part of the Monte
Carlo simulation of the $M(pK_S)$ spectrum, whereas the same peak generated in
the $M_{X}(K_S)$ spectrum can be clearly reconstructed.

\vskip 2cm

\begin{figure}[htb!]
\includegraphics[width=3.0in] {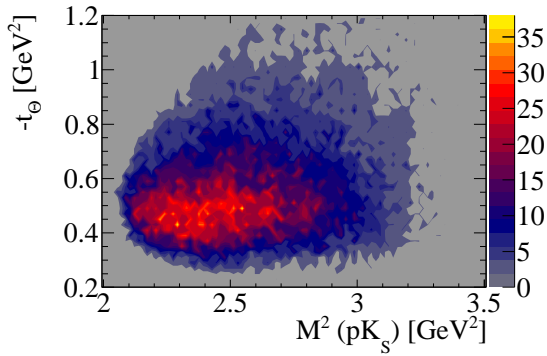}
\caption{(Color online). $-t_{\Theta}$ versus $M(pK_S)^2$ for events
  selected under the $\phi$ peak, $M_X(p)=1.02\pm0.01$~GeV.}
\label{fig:t_vs_mpks}
\end{figure}

In Fig.~\ref{fig:t_vs_mpks} we plot $-t_{\Theta}$ versus
$M(pK_S)^2$ to see if there is any correlation between these two
variables due to the limited CLAS acceptance, although these two
variables are in general independent. As one can see there is no
strong correlation just as expected.

\section{Monte Carlo Simulation and Statistical Significance of the Observed Structure }
 
In our analysis we looked for a possible resonance structure that interferes with
$\phi$-production in the final state $K_SK_Lp$. We looked for
deviation of the missing mass spectra of $K_S$ in the experimental
data from the missing mass of $K_S$ for pure $\phi$-production. 

Our $\phi$ photoproduction Monte-Carlo simulation is based on
the Titov-Lee model~\cite{titov}. The angular dependencies of
$\phi$-decay were taken from the pomeron exchange model and the energy
dependence of $\phi$-production was modeled using several iterations
in the simulation. The $t-$dependence of  $\phi$-production was
simulated with an exponential function $\exp(b_{\phi}t_{\phi})$, here
$t_{\phi}=(P_{\gamma}-P_{\phi})^2$, where the slope
$b_{\phi}=3.4$~GeV$^{-2}$ was taken
from the existing data~\cite{mibe}. The
model describes experimental
data quite well in the low $t_{\phi}$
region, where $\phi$ production due to the pomeron exchange mechanism dominates.

 Simulated events were passed through the CLAS detector emulation
 program (GSIM) and then were reconstructed with the RECSIS (CLAS 
 reconstruction program). The Monte-Carlo simulated data were analyzed
 using the same programs and the obtained distributions (with the same cuts as
 for the data) were compared to the missing mass of $K_S$ from 
 experimental data.
 
\begin{figure}[htb!]
\includegraphics[width=3.40in] {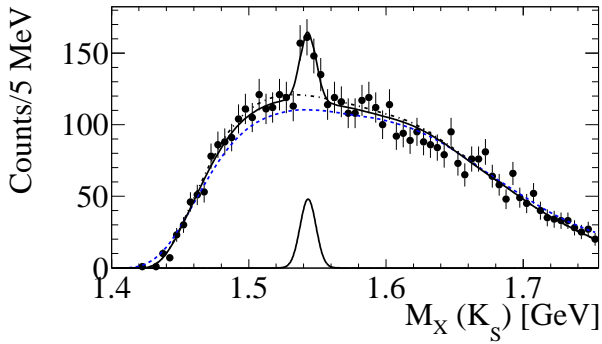}
\caption{Missing mass of $K_S$ with a cut
  $-t_{\Theta}<0.45$GeV$^2$. The dashed line is a result of $\phi$ MC simulation, the dashed-dotted line is a modified MC distribution, and the solid line is
  a result of the fit with modified MC distribution plus Gaussian function.}
\label{fig:mpkl_fin}
\end{figure}

 In Fig.~\ref{fig:mpkl_fin} the experimental distribution of the missing mass of $K_S$, $M_{X}(K_S)$, is presented with the cut on 
 $-t_{\Theta}<0.45$~GeV$^2$. 
 The dashed line is the result of the Monte Carlo simulation, which is a
 smooth distribution without any structure.
To account for imperfections in the detector simulation, we allowed this distribution to vary slightly to describe the data better.

 For this the missing mass distribution is fitted using the function:
\begin{equation}
F_{B} = SIM(\phi)\cdot POL_3, 
\label{eq:FIT_EQ}
 \end{equation}
\noindent where $SIM(\phi)$ is the  Monte Carlo  simulated histogram
from $\phi-$production, and $POL_3$ is a 3rd order polynomial
function. All parameters of the $POL_3$ function were allowed to vary. The result is the dashed-dotted line in Fig.~\ref{fig:mpkl_fin}; we refer to that
distribution ($F_{B}$) as the null or background (B) hypothesis,
i.e. assuming that experimental spectrum is fully described 
by the modified Monte Carlo distribution. 

A second hypothesis assumes
that, in addition to the background described by the null hypothesis,
there is a resonance structure, which is chosen to have Gaussian (G)
shape. 

This is called the signal+background hypothesis (S+B) and fit with the following function:
 \begin{equation}
F_{S+B} = SIM(\phi)\cdot POL_3 + G,
\label{eq:FIT_EQ}
 \end{equation}
 \noindent shown as the solid line in Fig.~\ref{fig:mpkl_fin}.
 
 To estimate the statistical significance of the observed resonance
 structure  we performed a log likelihood test of the two hypotheses:
 
 \begin{align}
-2\ln \mathcal {L}_{S+B} = 2\sum_{i=0}^{N}{\big [(s_i+b_i) - n_i + n_i\cdot \ln(n_i/(s_i+b_i))\big ] },\\
-2\ln \mathcal {L}_{B} = \sum_{i=0}^{N}{\big [b_i - n_i + n_i\cdot \ln (n_i/b_i)\big ]},
\end{align}
where $\mathcal {L}_{S+B}, \mathcal {L}_B$ are likelihoods for S+B and B hypotheses
respectively, $n_i$ is a total number of events in the $i$-th bin, $b_i$ and $s_i$ are
the number of predicted background and signal events
in the given bin. 
The peak parameters obtained from the fit are
$M_{X}(K_S)=1.543\pm 0.002$~GeV with a Gaussian width, $\sigma=0.006\pm
0.001$~GeV, compatible with the experimental resolution of CLAS~\cite{CLAS1}.

\begin{figure}[htb!]
\includegraphics[width=3.40in] {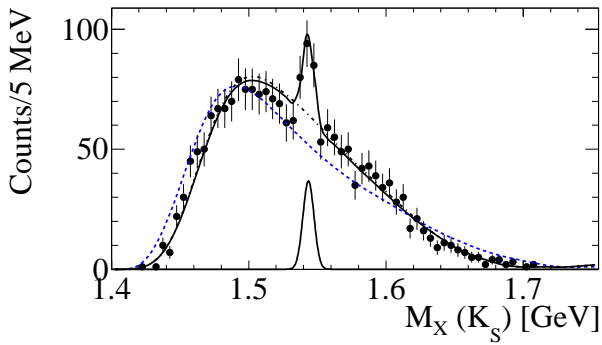}
\caption{Missing mass of $K_S$  with cuts: $-t_{\Theta}<0.45$~GeV$^2$
and $M(pK_S)<1.56$~GeV. The dashed line is a result of $\phi$ MC simulation, the dashed-dotted line is a modified MC distribution, and the solid line is
  a result of the fit with modified MC distribution plus Gaussian function.}
\label{fig:mpkl_fin_mpks}
\end{figure}

To demonstrate how well the Monte Carlo simulation reproduces the
shape of the experimental distribution and to see the robustness of
the significance of the observed signal, 
 we present the $M_{X}(K_S)$ distribution with cuts on
 $-t_{\Theta}<0.45$~GeV$^2$ and invariant mass $M(pK_S)<1.56$~GeV in  Fig.~\ref{fig:mpkl_fin_mpks}.
 The additional cut on the invariant mass $M(pK_S)$ changes the shape
 of the experimental distribution significantly. Now the resonance
 structure appears on top of a background with inclined shape and not
 in the middle of the symmetric distribution, as in Fig.~\ref{fig:mpkl_fin}. The fit values for the peak are
 $M_{X}(K_S)=1.543 \pm 0.001$~GeV with a Gaussian width, $\sigma=0.004\pm 0.001$~GeV.

\vskip 0.5cm

\begin{figure}[htb!]
\includegraphics[width=3.40in] {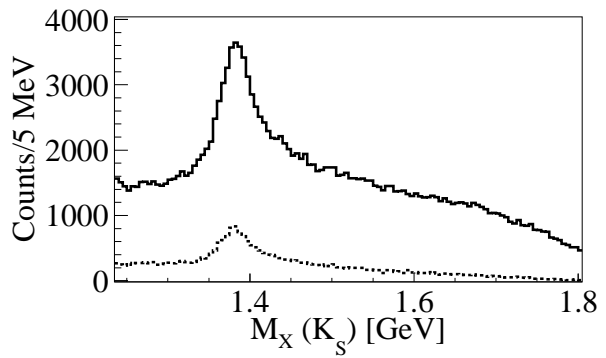}
\caption{Missing mass of $K_S$  for events with
  $|M_{X}(pK_S)-0.497|>0.03$~GeV. Solid histogram without any cut on
  $t_{\Sigma}$, and dashed histogram with a cut $-t_{\Sigma}<0.45$~GeV$^2$}
\label{fig:no_sigma}
\end{figure}

Below, in Table~\ref{tab:stat} we summarize statistical information about
hypotheses testing based on data presented in
Fig.~\ref{fig:mpkl_fin} and Fig.~\ref{fig:mpkl_fin_mpks} as a result
of the fits described above. For each of
these fugures there are two rows with the fit corresponding to S+B and
B hypotheses. The columns represent
number of degrees of freedom (ndf), $\chi^2$, p-value, and log
likelihood, $\ln \mathcal {L}$, for a given
hypothesis. The seventh column is two times the log likelihood ratio of
two hypotheses, $2\Delta(\ln \mathcal {L})$, a square root of which for one degree of
freedom difference between the two hypotheses would have normally represented
statistical significance in number of $\sigma$'s, however in our case with
$\Delta ndf=3$, it will be lower. To avoid obtaining artificially high
statistical signifcance we took into account the fact that
the hypothesis with the less degrees of freedom has an advantage to
fit data better. To do this we recalculated $\sqrt{2\Delta(\ln \mathcal {L})} =\sqrt{\chi^2_{(\Delta ndf=1)}}$ 
significance for one degree of freedom difference using p-value
corresponding to $\chi^2_{(\Delta ndf=3)}=2\Delta(\ln \mathcal { L})$ for three
degrees of freedom. The obtained significance $5.3\sigma$ for
Fig.~\ref{fig:mpkl_fin}, and $4\sigma$ for Fig.~\ref{fig:mpkl_fin_mpks} is presented in the column S, in units
of $\sigma$. The fitted signal yield (number of events), i.e. the integral of the Gaussian
distribution,  is presented in the last column with statistical errors from the fit.
\vskip 0.5cm
\begin {table} [h]
\caption {Fit Results (see text for explanation).}
\label {tab:stat}
\begin{center}
\begin{tabular}{|l|l|l|l|l|l|l|l|l|}
\hline
Figure & Fit& ndf & $\chi^2$ & p & $-2\ln {\mathcal{L}}$ &
$2\Delta(\ln {\mathcal L})$ &S & Signal Yield\\
\hline 
Fig.~\ref{fig:mpkl_fin}& S+B& 67 & 52 & 0.91 & 54  & 31& 5.3$\sigma$ &$142\pm46$\\
 &B& 70 & 79 & 0.22 & 85 & &  &\\
\hline 
 Fig.~\ref{fig:mpkl_fin_mpks} & S+B & 46 & 37 & 0.82 & 40 & 22& 4$\sigma$
 & 83$\pm$27\\
  &B& 49 & 55 & 0.24 & 62 & &  &\\ 
\hline
\end{tabular}
\end{center}
\end{table}
\vskip 0.5cm
As strangeness in the $pK_L$ system is not fixed, the structure
in Figs.~\ref{fig:mpkl_fin} and~\ref{fig:mpkl_fin_mpks} may be due to
an unobserved $\Sigma^{*+}$
resonance, which decays into $p\bar K^0$. However $\Sigma^{*+}$ should
also decay to $\Lambda\pi$, $\Lambda\pi\pi$, $\Sigma\pi$,
$\Sigma\pi\pi$, etc. In order to check this possibility, in
Fig.~\ref{fig:no_sigma} we plot the missing mass of $K_S$, $M_{X}(K_S)$,
for events outside the peak of the missing kaon,
$|M_{X}(pK_S)-0.497|>0.03$~GeV. As one can see there is a clear peak
of $\Sigma(1385)$, but no narrow resonance structure is
seen at $\sim$1.54~GeV neither without a cut on $t_{\Sigma}=(P_{\gamma}-P_{K_S})^2$ (solid histogram) nor with
a cut $-t_{\Sigma}<0.45$~GeV$^2$ (dashed histogram). The
Fig.~\ref{fig:no_sigma} demonstrates also that the set of cuts
 which we use in our analysis (vertex cuts and $t_{\Theta}$-cut) does not produce by itself  any
 artificial peaks.

\section{Conclusions}

To conclude, we use, for the first time, meson-baryon interference
to search for a weak baryon resonance in the same final state.
This search was
   motivated by a desire to increase the sensitivity of the CLAS
   detector to a possible pentaquark state.
We observe a narrow structure in the data at $ M_X(K_S)=
1.543$~GeV with a Gaussian width $\sigma = 0.006$~GeV for the reaction
$\gamma p\rightarrow p\,K_SK_L$ when $M(K_SK_L)=m_\phi$ and $-t_{\Theta} < 0.45$ GeV$^2$.  Because we are looking for a peak in the interference between
a resonance in the $KN$ system and $\phi$ production, all of our background is due to $\phi$ production. This puts us
in the advantageous position of understanding the background in our reaction. The peak is
not reproduced by the MC simulation that accurately describes the essential background of $\phi$
production.

The statistical significance of the observed signal, estimated as a log likelihood ratio
of signal+background and background-only hypotheses is $5.4\sigma$. When we vary
the background by cutting on the invariant mass $M(pK_S)$, the peak remains significant.

The best explanation for the observed structure is interference
between $\phi$ and $KN$  resonance production. 
Since strangeness is not fixed in this reaction, there are two
possibilities for the origin of the observed structure. It may be due
to the photoproduction of the $\Theta^+$ pentaquark or some 
unknown $\Sigma^*$ resonance. As we did not observe a narrow $\Sigma^{*+}$ 
 decaying to ground state $\Lambda$ and $\Sigma$ hyperons,
it is unlikely for the observed structure to be due to a $\Sigma^*$ resonance.
Note that the interference can shift the peak position from the actual resonance position.  To simulate in  detail the
interference between two subprocesses one needs to have much more
information, including the cross section and width of the baryon resonance, the slope of its
$t$-dependence and the relative phase of the interfering amplitudes. The existing data
set is too small to constrain reliably any of these parameters, so we leave such
studies for the future.

The present result does not contradict the previous CLAS analysis in
the same channel \cite{CLAS1} that did
not observe a peak near 1.54~GeV, 
since  events with
$M(K_SK_L)=m_\phi$ were excluded there with the cut $m_{X}(p)>1.04$~GeV. Assuming that the observed peak is mainly due to $\phi-KN$
interference, we estimated the photoproduction cross section of the $KN$ resonance with a Breit-Wigner width $\Gamma$=1~MeV to be two orders of magnitude smaller than the photoproduction cross section of the $\phi$ meson, consistent
with the few nanobarn upper limit of the cross section established by the CLAS collaboration~\cite{CLAS1}
for the photoproduction of $\Theta^+$.

In addition, because the peak in these data is only observed at
relatively small values of $t_\theta$, this might 
reconcile the CLAS null results~\cite{CLAS1} and the
SPring-8 observation of the $\Theta^+$~\cite{nakano2} in similar
channels. The CLAS acceptance at $-t_\Theta < 0.5$~GeV$^2$ is much
smaller than that of SPring-8.

Nevertheless, we are not without unanswered questions. One of those
questions is: if the observed signal is due to the interference with $\phi$ meson
production, why does the statistical significance of the signal sharply
diminish at higher values of $t_{\Theta}$? Is it because the phase of the
interference has a strong t-dependence or is it because the mechanism of $\phi$
production changes from pomeron 
exchange to the excitation of intermediate baryon resonances and
therefore an increase of statistics in 
$\phi$ production does not necessarily guarantee the same increase in the interference term?

Another question is: why do restrictions on the invariant mass of the
$pK_S$ system effectively enable this signal to manifest itself, even without the t-cut?
Is it possible that  well known excited $\Sigma^*$ resonances listed in~\cite{pdg} interfere destructively with the $\phi$
and affect the narrow structure we observe at 1.54 GeV?

To answer these questions, to further corroborate the existence of a resonance underlying the observed structure,
to elucidate its quantum numbers, and to understand the details of the interference,
additional data for this and other channels are needed. 

The interpretation of experimental results obtained in this analysis reflects
opinion of the authors and not that of the CLAS Collaboration as a whole.

We would like to acknowledge the outstanding efforts of the staff of the Accelerator and
the Physics Divisions at Jefferson Lab that made the experiment
possible. 
This work was
supported in part by the Italian Istituto Nazionale di Fisica Nucleare, the French Centre
National de la Recherche Scientifique and Commissariat \`a l'Energie Atomique, the U.S.
Department of Energy and National Science Foundation, and the Korea Science and Engineering
Foundation. The Southeastern Universities Research Association (SURA) operates the Thomas
Jefferson National Accelerator Facility for the United States Department of Energy under
contract DEAC05-84ER40150. The work of M.V.~Polyakov is supported by
the DFG SFB/TR16 Program (Germany). Ya.Azimov is partly supported by the Russian State grant RSGSS-65751.2010.2.



\begin{thebibliography}{99}
\bibitem{DIAKON}
D.~Diakonov, V.~Petrov, and M.~V.~Polyakov,
Z.\ Phys.\  A {\bf 359}, 305 (1997)

\bibitem{nakano} T.~Nakano {\it et. al.}, Phys. \ Rev. \ Lett. {\bf 91}, 012002 (2003).
\bibitem{DIANA} V.V.~Barmin {\it et al.}, [DIANA Collaboration], Phys. \ Atom. \ Nucl. {\bf 66}, 1715 (2003); Yad. \ Fiz. {\bf 66}, 1763 (2003).
\bibitem{CLASD} S.~Stepanyan {\it et al.}, [CLAS Collaboration], Phys. \ Rev. \ Lett. {\bf 91}, 252001 (2003).
\bibitem{CLASH} V.~Kubarovsky {\it et al.}, [CLAS Collaboration], Phys. \ Rev. \ Lett. {\bf 92}, 032001 (2004).
\bibitem{SAPHIR} J.~Barth {\it et al.}, [SAPHIR Collaboration], Phys. \ Lett. B {\bf 572}, 127 (2003).
\bibitem{HERMES} A.~Airapetian {\it et al.}, [HERMES Collaboration], Phys. \ Lett. B {\bf 585}, 213 (2004).
\bibitem{ZEUS} S.~Chekanov {\it et al.}, [ZEUS Collaboration], Phys. \ Lett. B {\bf 591}, 7 (2004).
\bibitem{SVD} A.~Aleev {\it et al.}, [SVD Collaboration], Yad. \
 Fiz. {\bf 68}, 1012 (2005).
\bibitem{HERAB} I.~Abt {\it et al.}, [HERA-B Collaboration], Phys. \
Rev. \ Lett. {\bf 93}, 212003 (2004). 
\bibitem{HyperCP} M.J.~Longo {\it et al.}, [HyperCP Collaboration], Phys. \
Rev. D {\bf 70}, 111101 (2004). 
\bibitem{BES} J.Z.~Bai {\it et al.}, [BES Collaboration], Phys. \
Rev. D {\bf 70}, 034506 (2004).  
\bibitem{ALEPH} R.~Barate {\it et al.}, [ALEPH Collaboration], Phys. \ Lett. {\bf 599}, 1 (2004).
\bibitem{BABAR} B.~Aubert {\it et al.}, [BABAR Collaboration], Phys. \
Rev. \ Lett. {\bf 95}, 042002 (2005).
\bibitem{CLASg10} B.~McKinnon {\it et al.}, [CLAS Collaboration], Phys. Rev. Lett. {\bf 96}, 212001 (2006).
 \bibitem{CLAS1}  R.~De Vita {\it et al.},  [CLAS Collaboration],  Phys.\ Rev.\  D {\bf 74}, 032001 (2006)
\bibitem{pdg}
K. ~Nakamura et al. (Particle Data
Group), J. Phys. G 37, 075021 (2010) 
\bibitem{hicks} K.~Hicks, Progr. \ Part. \ Nucl. \ Phys. {\bf 55}, 647 (2005)
\bibitem{burkert}
V.D.~Burkert, Int. \ J. \ Mod. \ Phys. \ A {\bf 21}, 1764 (2006)
\bibitem{AGS} Y.~Azimov, K.~Goeke, and I.~Strakovsky, Phys. \ Rev. D {\bf 76}, 074013 (2007)
\bibitem{nakano2} T.~Nakano {\it et. al.}, Phys. \ Rev. \ C {\bf 79}, 025210 (2009)
\bibitem{diana2}  V.V.~Barmin {\it et al.}, [DIANA Collaboration], Phys. \ Atom. \ Nucl. {\bf 70}, 35 (2007);
\bibitem{diana3}  V.V.~Barmin {\it et al.}, [DIANA Collaboration],
 Phys. \ Atom. \ Nucl. {\bf 73}, 1168 (2010);
\bibitem{azimov} Ya.~Azimov, J. \ Phys. \ G: \ Nucl. \ Part. \ Phys. {\bf 37}, 023001 (2010) 023001
\bibitem{amarian}  M.~Amarian, D.~Diakonov, and M.~V.~Polyakov,
Phys.\ Rev.\  D {\bf 78}, 074003 (2008).
\bibitem{mecking} B.~Mecking et al., Nucl. Inst. and Meth. A {\bf 503},
513 (2003).
\bibitem{williams} R.A.~Williams, Phys.\ Rev.\  C {\bf 57}, 223 (1998).
\bibitem{titov} A.~I.~Titov and T.~S.~H.~Lee, Phys.\ Rev.\  C {\bf 67}, 065205~(2003).
\bibitem{mibe} T.~Mibe {\it et. al.}, [LEPS Collaboration], Phys. \ Rev. \ Lett. \ {\bf 95}, 182001 (2005).



\end{thebibliography}
\end{document}